\documentclass[pra,showpacs,twocolumn]{revtex4}
\usepackage{dcolumn}
\usepackage{bm}
\usepackage{amssymb}
\usepackage{amsmath}
\usepackage{amsfonts}

\usepackage[]{graphicx}
\begin{document}
\bibliographystyle{prsty}
\title{Equivalence between the Hamiltonian and Langevin noise description of plasmon-polaritons in a dispersive and lossy inhomogeneous medium}
\author{ Aur\'elien Drezet $^{1}$}
\address{(1) Univ.~Grenoble Alpes, CNRS, Institut N\'{e}el, F-38000 Grenoble, France}
\begin{abstract}
We demonstrate the fundamental links existing between two different descriptions of quantum electrodynamics in inhomogeneous, lossy and dispersive  dielectric media which are based either on the Huttner-Barnett formalism for polaritons [B. Huttner and S. M. Barnett, Phys.Rev. A \textbf{46}, 4306 (1992)] or the Langevin noise approach using fluctuating currents [T. Gruner and D.-G. Welsch, Phys.Rev.A \textbf{53}, 1818 (1996)]. In this work we demonstrate  the practical equivalence of the two descriptions by introducing the concept of effective  photon state associated with some specific noise current distribution. We study the impact of these results on the  calculation  and interpretation of quantum observables such as fluctuations, correlations, and Casimir forces.         
\end{abstract}

 \maketitle
\section{Introduction}
\indent The recent advances in quantum electrodynamics (QED) at the nanoscale in a metallic environment~\cite{Novotny,Barnes}, i.e., quantum-nano-plasmonics (QNP) \cite{Tame2013,Agio},  opened up many possibilities for integrated quantum technologies. 
One of the central issue in this field is the control over the coupling between fluorescent quantum emitters and nano-antennas or compact plasmonic devices~\cite{Tame2013,Agio}. Experimentally, many key results have been obtained in the last decade thank to the development of methods like active probe near-field optical microscopy~\cite{opex2006,micron}, and to rapid progress in nanofabrication technics and particle synthesis~\cite{Colas2,Chicanne,Colas3,Cuche2010,Pham,Martin}.\\
\indent From the theoretical point of view however, one of the most challenging issue is still to propose a rigorous quantized formalism for QNP including the intrinsic dispersion and dissipation of metallic inhomogeneous systems. Over the years two different strategies have been proposed to tackle this difficult problem. The first general approach~\cite{Huttner1991,Huttner1992a,Huttner1992b,Huttner1992c,Matloob1995,Matloob1996,Barnett1995} is based on the canonical quantization by Huttner and Barnett  of the Hamiltonian describing  the coupling between light and dielectric matter that includes a bath of oscillators to model the dispersive and dissipative properties of the surrounding medium. This approach, extending the seminal works of Hopfield and Fano for polaritons~\cite{Fano1956,Hopfield1958,Huang1951}, was established rigorously only for the homogeneous medium case. However, its generalization to structured nano-systems lacked for many years. Therefore, a second  more powerful strategy, based on a  dipolar Langevin noise (DLN) formalism~\cite{Gruner1995,Gruner1996,Yeung1996}, was favored in which no canonical foundation was required.  Instead, fluctuating currents are phenomenologically added to deal with the problem of dissipation and dispersion. This approach was intensively used in the literature~\cite{Scheel1998,Dung1998,Dung2000,Scheel2001,Matloob1999,Matloob2004,Fermani2006,Raabe2007,Amooshahi2008,Scheelreview2008}, e.~g., for describing optical Bloch equations in the weak or strong optical coupling in QNP~\cite{Dzotjan2010,Cano2011,Hummer2013,Chen2013,Delga2014,Hakami2014,Choquette2012,Grimsmo2013,Rousseaux2016}, Casimir interactions, quantum frictions and thermal fluctuating forces~\cite{Tomas2002,Buhman2004,Intravaia2014,Intravaia2016,PhilbinCasimir}, and more recently for modeling quantum optical non-linearities such as spontaneous down conversion of photon pairs~\cite{Scheel2006,Poddubny2016}. It is central to observe that the DLN approach is a direct development of the historical works by Rytov and others~\cite{Lifshitz1956,Ginzburg,Rytov,Milonnibook} which, based on some considerations about the standard fluctuation dissipation theorem for electric currents~\cite{Callen1951}, was used for justifying Casimir and thermal forces (for recent developments of such phenomenological `fluctuational electrodynamics' techniques in the context of nanotechnology see \cite{Rosa2010,Agarwal1975,Sipe1984,Rousseau,Henkel,Otey2014}). Few years ago, it was proposed that the equivalence between the Hamiltonian and DLN approaches should finally be rigorous~\cite{Wubs2001,Suttorp2004a,Suttorp2004b,Suttorp2007,Bhat2006,Judge2013,Philbin2010}. However, we recently showed \cite{A,B,C} that a full Hamiltonian description, generalizing the Huttner-Barnett results ~\cite{Huttner1991,Huttner1992a,Huttner1992b,Huttner1992c,Matloob1995,Matloob1996,Barnett1995} and valid for any inhomogeneous dielectric systems, must not only include  the material oscillator degrees of freedom, i.e., like in the DLN method, but also add the previously omitted quantized photonic degrees of freedom associated with fluctuating optical waves coming from infinity and scattered by the inhomogeneities of the medium~\cite{B}. Furthermore, the inclusion of both photonic and material fluctuations on a equal footing appears necessary in order to preserve the full unitarity of the quantum evolution and to conserve time symmetry\cite{Milonnibook}.\\
\indent However, from a pragmatic perspective it is still crucial to understand why the DLN approach works so well and to justify its foundation on a solid ground.   Here, we will present such a demonstration and show how to justify for all practical purposes of  QNP the application of DLN methods, i.e., by removing the independent photonic degrees of freedom though without breaking unitarity and time symmetry.\\
\indent The layout of this work is as follows:  In Sec. II we give a summary of the main ingredients associated with the generalized Huttner-Barnett approach and the DLN method and stress the similarities and differences. In Sec. III we give a demonstration of the equivalence between the two approaches by defining a new effective medium located at spatial infinity. In particular we show that we must include in the DLN an effective pure photon field which has all the classical and quantum properties of a free photon state scattered by a dielectric system. In Sec. IV we analyze some consequences of our finding for the effective calculations and physical interpretations of QNP observables such as local density of states (LDOS), quantum fluctuations and correlations, and Casimir and thermal forces. We conclude with a summary and some perspectives in Sec. V.     
\section{The General Hamiltonian for the description of a lossy dielectric medium}
\subsection{Contribution of photonic and material degrees of freedom to the electric field operator }
\indent We start with the canonical description given in \cite{A,B} in the Heisenberg picture. It is  based on a dual formalism involving an electric potential vector operator $\textbf{F}(\textbf{x},t)$ such that $\boldsymbol{\nabla}\cdot\textbf{F}=0$ (dual Coulomb gauge) and $\textbf{D}=\textbf{E}+\textbf{P}=\boldsymbol{\nabla}\times \textbf{F}$, where $\textbf{D}$ is the transverse displacement field, $\textbf{E}$ the electric field, and  $\textbf{P}$ the total dipole density of the medium. $\textbf{P}$ is the sum of the induced dipole density $\int_{0}^{t-t_0}\chi(\mathbf{x},\tau)d\tau\mathbf{E}(\mathbf{x},t-\tau)$, characterized by the initial time $t_0$ and the linear dielectric susceptibility $\chi(\mathbf{x},\tau)$ (i.e., satisfying Kramers-Kr\"{o}nig relations), and $\mathbf{P}^{(0)}(\mathbf{x},t)$ the fluctuating dipole density given by 
$\int_0^{+\infty}d\omega\sqrt{\frac{\hbar\varepsilon''_{\omega}(\mathbf{x})}{\pi}}[\mathbf{f}^{(0)}_{\omega}(\mathbf{x},t)+\mathbf{f}^{\dagger(0)}_{\omega}(\mathbf{x},t)]$, with $\varepsilon''_{\omega}:=\textrm{Imag}[\varepsilon_{\omega}]$ the imaginary part of the local dielectric permittivity $\varepsilon_{\omega}=\varepsilon'_{\omega}(\mathbf{x})+i\varepsilon''_{\omega}$. In this description $\mathbf{f}^{(0)}_{\omega}$ and $\mathbf{f}^{\dagger(0)}_{\omega}$ are respectively lowering and rising bosonic vector field operators associated with the fluctuating bath of material oscillators, i.e., rigorously equivalent to those operators given in the standard DLN approach. Moreover, in \cite{A,B,C} we showed that these noise operators are related to the total field operators at the initial time $t_0$, i.e., $\mathbf{f}^{(0)}_{\omega}(\mathbf{x},t)=\mathbf{f}_{\omega}(\mathbf{x},t_0)e^{-i\omega(t-t_0)}$. This is essential since the choice of retarded causal Green functions involves necessarily a boundary condition in the remote past at $t_0<t$. Therefore as discussed in \cite{B} our formalism preserves time symmetry and allows other equivalent descriptions involving `advanced' Green functions and boundary conditions at a future time $t_f>t$. The present choice is of course dictated by physical considerations not part of QED but connected to thermodynamics and cosmology. We also point out that in the general case, i.e.,  when external systems such as  fluorescent molecules are coupled to the fields we have to add  to $\textbf{P}$ a contribution $\textbf{P}^{(\textrm{mol.})}(\mathbf{x},t)$ \cite{C} which we let here unspecified. Within this dual formalism we can show that the electromagnetic field  operators satisfy Maxwell's equations and, since both $\textbf{D}$ and the magnetic field $\textbf{B}$ are transverse, it is not necessary to make the distinction between transverse  quantized and longitudinal otherwise un-quantized fields~\cite{A}.\\ 
\indent Using a Lagrangian or Hamiltonian description we obtain a formal separation of the electric field as:  $\mathbf{E}(\mathbf{x},t)=\mathbf{E}_{\textrm{in}}^{(v)}(\mathbf{x},t)+\mathbf{E}_{\textrm{ret.}}^{(v)}(\mathbf{x},t)$  where  $\mathbf{E}_{\textrm{in}}^{(v)}(\mathbf{x},t)$ is the incident field associated with pure propagative photons while the second term $\mathbf{E}_{\textrm{ret.}}^{(v)}(\mathbf{x},t)$ corresponds to the total scattered field induced by $\mathbf{P}$ which depends on the Green dyadic propagator $\boldsymbol{\Delta}_{\textrm{ret.}}^{(v)}(\tau,\mathbf{x},\mathbf{x'})$ in vacuum \cite{A,B}. We have explicitly  \begin{eqnarray}
\mathbf{E}_{\textrm{ret.}}^{(v)}(\mathbf{x},t)=\int_0^{t-t_0}d\tau\int d^3\mathbf{x'} \boldsymbol{\Delta}_{\textrm{ret.}}^{(v)}(\tau,\mathbf{x},\mathbf{x'}) \cdot\mathbf{P}(\mathbf{x'},t-\tau). \nonumber\\ \label{0}
\end{eqnarray}
Writing the Fourier expansion of the electric field $\widetilde{\mathbf{E}}_\omega(\mathbf{x})=\int_{-\infty}^{+\infty}\frac{dt}{2\pi} \mathbf{E}(\mathbf{x},t)e^{+i\omega t} $, i.e., with $t_0\rightarrow-\infty$ [in \cite{B} we used instead the forward Laplace's transforms which works for arbitrary $t_0$] we have 
\begin{eqnarray}
\widetilde{\mathbf{E}}_{\omega}(\mathbf{x})=\widetilde{\mathbf{E}}_{\textrm{in},\omega}^{(v)}(\mathbf{x})+\int d^3 \mathbf{x'}\frac{\omega^2}{c^2}\mathbf{G}_\omega^{(v)}(\mathbf{x},\mathbf{x'})\cdot\widetilde{\mathbf{P}}_\omega(\mathbf{x'}) \label{1}
\end{eqnarray}                                       
where $\mathbf{G}_\omega^{(v)}(\mathbf{x},\mathbf{x'})$ is the stationary and retarded dyadic Green function  solution of $\boldsymbol{\nabla}\times\boldsymbol{\nabla}\times\mathbf{G}_\omega^{(v)}(\mathbf{x},\mathbf{x'})-\frac{\omega^2}{c^2}\mathbf{G}_\omega^{(v)}(\mathbf{x},\mathbf{x'})=\mathbf{I}\delta(\mathbf{x}-\mathbf{x'})$ and such that $\boldsymbol{\Delta}_{ret}^{(v)}(\tau,\mathbf{x},\mathbf{x'}) =\int_{-\infty}^{+\infty}\frac{d\omega}{2\pi} e^{-i\omega\tau} \frac{\omega^2}{c^2}\mathbf{G}_\omega^{(v)}(\mathbf{x},\mathbf{x'})$. The free-field $\widetilde{\mathbf{E}}_{\textrm{in},\omega}^{(v)}(\mathbf{x})$ is expanded into plane-waves of pulsations $\omega_\alpha$ such as 
\begin{eqnarray}
\tilde{\mathbf{E}}_{\textrm{in},\omega}^{(v)}(\mathbf{x})=\sum_{\alpha,j}[\mathbf{E}_{\alpha,j}^{(v)}(\mathbf{x})c_{\alpha,j}^{(v)}(t_0)e^{i\omega_\alpha t_0}\delta(\omega-\omega_\alpha)\nonumber\\+\mathbf{E}_{\alpha,j}^{\ast(v)}(\mathbf{x})c_{\alpha,j}^{\dagger(v)}(t_0)e^{-i\omega_\alpha t_0}\delta(\omega+\omega_\alpha)]\label{2}\end{eqnarray} 
 where $c_{\alpha,j}^{(v)}$, and $c_{\alpha,j}^{\dagger(v)}$ are respectively the lowering and rising vacuum photon operators satisfying usual commutation relations for bosons and associated with the plane wave modes $\mathbf{E}_{\alpha,j}^{(v)}(\mathbf{x})$ (i.e., labeled by the quantized wave-vector $\mathbf{k}_\alpha$ and the transverse  polarization $\boldsymbol{\hat{\epsilon}}_{\alpha,j}$, with $j=1,2$ \cite{A,B}) which are forming a complete orthogonal basis (with $\int d^3 \mathbf{x}\mathbf{E}_{\alpha,j}^{(v)}(\mathbf{x})\cdot\mathbf{E}_{\alpha',j'}^{(v)\ast}(\mathbf{x})=\frac{\hbar \omega_\alpha}{2}\delta_{\alpha,\alpha'}\delta_{j,j'}$, $\omega_\alpha=c|\mathbf{k}_\alpha|$) in agreement with Born-von Karman  periodic boundary conditions in a large rectangular box of volume $V_{\textrm{BK}}\rightarrow+\infty$~\cite{A,B}.\\
\indent  Morever, while Eq.~\ref{0} corresponds to a microscopic description, in macroscopic QED, e.g., in QNP, it is more convenient to consider a different separation of the electric field reading 
\begin{eqnarray}
\mathbf{E}(\mathbf{x},t)=\mathbf{E}_{\textrm{in}}^{(\textrm{eff.})}(\mathbf{x},t)+\mathbf{E}_{\textrm{ret.}}^{(\textrm{eff.})}(\mathbf{x},t)=\mathbf{E}_{\textrm{in}}^{(\textrm{eff.})}(\mathbf{x},t)\nonumber\\+\int_0^{t-t_0}d\tau\int d^3\mathbf{x'} \boldsymbol{\Delta}_{\textrm{ret.}}^{(\textrm{eff.})}(\tau,\mathbf{x},\mathbf{x'}) \cdot\mathbf{P}^{(\textrm{eff.})}(\mathbf{x'},t-\tau) 
 \label{3}
\end{eqnarray} where $\mathbf{E}_{\textrm{in}}^{(\textrm{eff.})}$ corresponds to the effective electromagnetic 'free field' solution of Maxwell's equations in the dielectric medium and where $\mathbf{E}_{\textrm{ret.}}^{(\textrm{eff.})}$ is the scattered field induced by the effective dipole distribution $\mathbf{P}^{(\textrm{eff.})}=\mathbf{P}^{(0)}+\textbf{P}^{(\textrm{mol.})}$ in presence of the dielectric. The retarded Green dyadic propagator in presence of the dielectric~\cite{B,C} $\boldsymbol{\Delta}_{ret}^{(\textrm{eff.})}(\tau,\mathbf{x},\mathbf{x'})$ is related to the usual time-independent effective Green tensor $\mathbf{G}_\omega^{(\textrm{eff.})}(\mathbf{x},\mathbf{x'})$ (i.e., $\boldsymbol{\Delta}_{ret}^{(\textrm{eff.})}(\tau,\mathbf{x},\mathbf{x'})=\int_{-\infty}^{+\infty}\frac{d\omega}{2\pi} e^{-i\omega\tau} \frac{\omega^2}{c^2}\mathbf{G}_\omega^{(\textrm{eff.})}(\mathbf{x},\mathbf{x'})$) which is solution of 
\begin{eqnarray}
\boldsymbol{\nabla}\times\boldsymbol{\nabla}\times\mathbf{G}_\omega^{(\textrm{eff.})}(\mathbf{x},\mathbf{x'})-\frac{\omega^2\varepsilon_{\omega}(\mathbf{x})}{c^2}\mathbf{G}_\omega^{(\textrm{eff.})}(\mathbf{x},\mathbf{x'})\nonumber\\=\mathbf{I}\delta(\mathbf{x}-\mathbf{x'}).
\label{Green}
\end{eqnarray}
 Writing once again the Fourier expansion of the electric field we have  $\widetilde{\mathbf{E}}_\omega(\mathbf{x})=\widetilde{\mathbf{E}}_{\textrm{in},\omega}^{(\textrm{eff.})}(\mathbf{x})+\widetilde{\mathbf{E}}_{\textrm{ret.},\omega}^{(\textrm{eff.})}(\mathbf{x})$ with 
\begin{eqnarray}
\widetilde{\mathbf{E}}_{\textrm{ret.},\omega}^{(\textrm{eff.})}(\mathbf{x})=\int d^3 \mathbf{x'}\frac{\omega^2}{c^2}\mathbf{G}_\omega^{(\textrm{eff.})}(\mathbf{x},\mathbf{x'})\cdot\widetilde{\mathbf{P}}^{(\textrm{eff.})}_\omega(\mathbf{x'}) \label{4}
\end{eqnarray} and  $\mathbf{G}_\omega^{(\textrm{eff.})}(\mathbf{x},\mathbf{x'})$ obeys the recursive Lippman-Schwinger relation:   
\begin{eqnarray}
\mathbf{G}_\omega^{(\textrm{eff.})}(\mathbf{x},\mathbf{x'})=\mathbf{G}_\omega^{(v)}(\mathbf{x},\mathbf{x'})+\int d^3 \mathbf{u}\frac{\omega^2}{c^2}\mathbf{G}_\omega^{(v)}(\mathbf{x},\mathbf{u})\nonumber\\ 
\cdot(\varepsilon_{\omega}(\mathbf{u})-1)\mathbf{G}_\omega^{(\textrm{eff.})}(\mathbf{u},\mathbf{x'})= \mathbf{G}_\omega^{(v)}(\mathbf{x},\mathbf{x'})\nonumber\\ 
+\int d^3 \mathbf{u}\frac{\omega^2}{c^2}\mathbf{G}_\omega^{(\textrm{eff.})}(\mathbf{x},\mathbf{u})
\cdot(\varepsilon_{\omega}(\mathbf{u})-1)\mathbf{G}_\omega^{(v)}(\mathbf{u},\mathbf{x'}). \label{5}
\end{eqnarray} Like for Eq.~\ref{2} the Fourier field  $\widetilde{\mathbf{E}}_{\textrm{in},\omega}^{(\textrm{eff.})}(\mathbf{x})$ is defined by 
\begin{eqnarray}
\widetilde{\mathbf{E}}_{\textrm{in},\omega}^{(\textrm{eff.})}(\mathbf{x})=\sum_{\alpha,j}[\mathbf{E}_{\alpha,j}^{(\textrm{eff.})}(\mathbf{x})c_{\alpha,j}^{(v)}(t_0)e^{i\omega_\alpha t_0}\delta(\omega-\omega_\alpha)\nonumber\\+\mathbf{E}_{\alpha,j}^{\ast(\textrm{eff.})}(\mathbf{x}) c_{\alpha,j}^{\dagger(v)}(t_0)e^{-i\omega_\alpha t_0}\delta(\omega+\omega_\alpha)\label{6}
\end{eqnarray}
 where $\mathbf{E}_{\alpha,j}^{(\textrm{eff.})}(\mathbf{x})$ are the classical electric fields which are solutions of the scattering problem of a plane wave $\mathbf{E}_{\alpha,j}^{(v)}(\mathbf{x})$ with pulsation $\omega_\alpha$ by the inhomogeneous dielectric medium~\cite{B,C}.
For these fields we have again the recursive Lippman-Schwinger relation~\cite{B,C}:
\begin{eqnarray}
\mathbf{E}^{(\textrm{eff.})}_{\alpha,j}(\mathbf{x})=\mathbf{E}_{\alpha,j}^{(v)}(\mathbf{x})+\int d^3 \mathbf{u}\frac{\omega_\alpha^2}{c^2}\mathbf{G}_{\omega_\alpha}^{(v)}(\mathbf{x},\mathbf{u})\nonumber\\ 
\cdot(\varepsilon_{\omega_\alpha}(\mathbf{u})-1)\mathbf{E}_{\alpha,j}^{(\textrm{eff.})}(\mathbf{u})=\mathbf{E}_{\alpha,j}^{(v)}(\mathbf{x})\nonumber\\
+\int d^3 \mathbf{u}\frac{\omega_\alpha^2}{c^2}\mathbf{G}_{\omega_\alpha}^{(\textrm{eff.})}(\mathbf{x},\mathbf{u})\cdot(\varepsilon_{\omega_\alpha}(\mathbf{u})-1)\mathbf{E}_{\alpha,j}^{(v)}(\mathbf{u})\label{7}
\end{eqnarray} 
 which results from the definition \cite{B}
\begin{eqnarray}
\widetilde{\mathbf{E}}_{\textrm{in},\omega}^{(\textrm{eff.})}(\mathbf{x})=\widetilde{\mathbf{E}}_{\textrm{in},\omega}^{(v)}(\mathbf{x})+\int d^3 \mathbf{u}\frac{\omega^2}{c^2}\mathbf{G}_{\omega}^{(v)}(\mathbf{x},\mathbf{u})\nonumber\\ 
\cdot(\varepsilon_{\omega}(\mathbf{u})-1)\widetilde{\mathbf{E}}_{\textrm{in},\omega}^{(\textrm{eff.})}(\mathbf{u})=\widetilde{\mathbf{E}}_{\textrm{in},\omega}^{(v)}(\mathbf{x})\nonumber\\
+\int d^3 \mathbf{u}\frac{\omega^2}{c^2}\mathbf{G}_{\omega}^{(\textrm{eff.})}(\mathbf{x},\mathbf{u})\cdot(\varepsilon_{\omega}(\mathbf{u})-1)\widetilde{\mathbf{E}}_{\textrm{in},\omega}^{(v)}(\mathbf{u}).\label{7b}
\end{eqnarray}
 \indent Importantly, contrarily to what occurred for the modal functions $\mathbf{E}_{\alpha,j}^{(v)}(\mathbf{x})$  the set of all the fields $\mathbf{E}^{(\textrm{eff.})}_{\alpha,j}(\mathbf{x})$ does not constitute in general an orthogonal basis of modes. Still,  $\widetilde{\mathbf{E}}_{\textrm{in},\omega}^{(\textrm{eff.})}(\mathbf{x})$ is completely determined by the knowledge of the operators $c_{\alpha,j}^{(v)},c_{\alpha,j}^{\dagger(v)}$ acting on genuine free-space photon states.\\
\begin{figure}[]
\centering
\includegraphics[width=8.5cm]{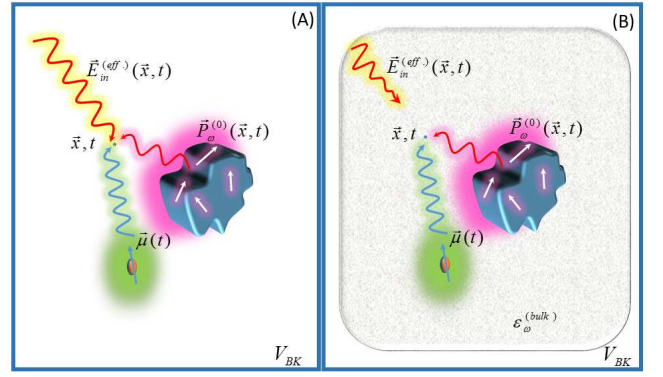}
\caption{Sketch of the two main approaches considered in the literature for modeling the coupling between quantum dipoles (here $\boldsymbol{\mu}(t)$) and any complex dielectric inhomogeneous environment. (A) shows the typical situation in the Huttner-Barnett formalism where the free space photon electric field  scattered by the environment $\mathbf{E}_{\textrm{in}}^{(\textrm{eff.})}(\mathbf{x},t)$ adds to the field produced  by the dipole $\boldsymbol{\mu}(t)$ and the dielectric fluctuating dipole distribution $\textbf{P}^{(0)}(\mathbf{x},t)$. (B) In the dipolar Langevin noise approach (DLN) the photon field is missing since it is absorbed by a residual bulk permittivity $\varepsilon_{\omega}^{(\textrm{bulk})}$ filling the complete Born von Karman quantization volume $V_{\textrm{BK}}$.} \label{figure1}
\end{figure}
\indent The present effective description adapted to QNP  (see Fig.~1(A)) considers on a equal footing the contributions from $\mathbf{E}_{\textrm{in}}^{(\textrm{eff.})}(\mathbf{x},t)$  and $\mathbf{E}_{\textrm{ret.}}^{(\textrm{eff.})}(\mathbf{x},t)$ and in \cite{B,C} we emphasized that both are necessary for preserving time symmetry and unitarity. Moreover, the arbitrariness concerning the time $t_0$ allows us to obtain other equivalent field separations, e.g., in which a contribution from the anti-causal permittivity  $\varepsilon^\ast_{\omega}(\mathbf{x})$ is included together with scattered waves depending on a boundary condition in the remote future at time $t_f$~\cite{B}.\\
\subsection{The Langevin Noise method seen from an Hamiltonian perspective}
\indent A particularly interesting and fundamental case concerns the homogeneous bulk medium with position independent permittivity $\varepsilon_{\omega}(\mathbf{x})$. Using the Laplace transform method we showed \cite{B} that $\mathbf{E}_{\textrm{in}}^{(\textrm{eff.})}(\mathbf{x},t)$  is exponentially damped in the future direction and is therefore vanishing at any point and any finite time in the limit $t_0\rightarrow -\infty$. Actually, rigorously speaking in classical physics where $c_{\alpha,j}^{(v)}$ are c-numbers and not operators (or q-numbers)  we can still obtain a finite value of the field  $\mathbf{E}_{\textrm{in}}^{(\textrm{eff.})}(\mathbf{x},t)$ in the bulk medium if some of the  $c_{\alpha,j}^{(v)}$ (defined at the initial time $t_0$~\cite{A,B}) are infinitely large. In  QED this is not possible but the initial state $|\Psi(t_0)\rangle$ can still be chosen in order to obtain infinite amplitudes at $t_0$ and therefore finite values at time $t\gg t_0$. Of course, the same is possible  in the anticausal representation using a finite time boundary condition at $t_f$ and which involves a field $\mathbf{E}_{\textrm{out}}^{(\textrm{eff.})}(\mathbf{x},t)$ exponentially growing in the future direction [in this alternative description the retarded and causal field $\mathbf{E}_{\textrm{ret.}}^{(\textrm{eff.})}(\mathbf{x},t)$ is replaced by an advanced and anticausal field  $\mathbf{E}_{\textrm{adv.}}^{(\textrm{eff.})}(\mathbf{x},t)$ \cite{B}]. Like before, the contribution of  $\mathbf{E}_{\textrm{out}}^{(\textrm{eff.})}(\mathbf{x},t)$ will not vanish if we impose specific boundary conditions $|\Psi(t_f)\rangle$ at  time $t_f$. At a fundamental level the problem is therefore perfectly symmetric. However, imposing infinite amplitudes in the past or in future to preserve time symmetry is not physically satisfactory and  this particular case occurs only because the infinitely extended bulk medium  (in space or time) is rather unphysical. Therefore, in order to remove these unwanted features of the model one can either suppose that the medium is not homogeneous for all time (for example before $t_0$ or after $t_f$). We can alternatively consider that the system is spatially very large compared to the relevant physical dimensions so that all the photonic components coming from infinity into the region of interest  (where $\varepsilon_{\omega}(\mathbf{x})\simeq$ Const.) are sufficiently damped ,i.e., $\mathbf{E}_{\textrm{in}}^{(\textrm{eff.})}(\mathbf{x},t)\rightarrow 0$, for all practical needs.   \\
\indent  The introduction of such an homogeneous lossy medium is intuitively associated with the DLN method. Indeed, in this approach the aim is to remove from the beginning the field $\mathbf{E}_{\textrm{in}}^{(\textrm{eff.})}(\mathbf{x},t)$. For this purpose Gruner and Welsch~\cite{Gruner1995,Gruner1996,Yeung1996}, and most authors after them, considered that by immersing any physical dipolar distribution $\mathbf{P}^{(\textrm{eff.})}=\mathbf{P}^{(0)}+\textbf{P}^{(\textrm{mol.})}$ and its associated inhomogeneous dielectric system  with local permittivity  $\varepsilon_{\omega}(\mathbf{x})$ into a infinitely extended bulk medium with causal permittivity $\varepsilon_{\omega}^{(\textrm{bulk})}$ they could ultimately give a clean Hamiltonian foundation to the DLN approach. In this strategy $\varepsilon_{\omega}^{(\textrm{bulk})}$ is supposed to be very close from  vacuum, i.e., $\varepsilon_{\omega}^{(\textrm{bulk})}\rightarrow 1+i0^+$ and thus should asymptotically lead to the ideal Langevin noise approach without photon field $\mathbf{E}_{\textrm{in}}^{(\textrm{eff.})}(\mathbf{x},t)\rightarrow  0$. In turn for the finite sources $\mathbf{P}^{(\textrm{eff.})}(\textbf{x},t)$ located in or near the inhomogeneities  the Green tensor $\mathbf{G}_{\omega}^{(\textrm{eff.})}(\mathbf{x},\mathbf{u})$ is assumed to be very close from the Green tensor in absence of the weakly dissipative  bulk medium  (i.e., with $\varepsilon_{\omega}^{(\textrm{bulk})}=1$). Therefore, the main postulate of the DLN formalism  (see Fig.~1(B)) is to write for the total electric field 
\begin{eqnarray}
\mathbf{E}(\mathbf{x},t)=\mathbf{E}_{\textrm{ret.}}^{(\textrm{eff.})}(\mathbf{x},t)\nonumber\\=\int_0^{t-t_0}d\tau\int d^3\mathbf{x'} \boldsymbol{\Delta}_{\textrm{ret.}}^{(\textrm{eff.})}(\tau,\mathbf{x},\mathbf{x'}) \cdot\mathbf{P}^{(\textrm{eff.})}(\mathbf{x'},t-\tau) 
 \label{3new}
\end{eqnarray}  where  the local permittivity $\varepsilon_{\omega}(\mathbf{x})$ is supposed identical to the one considered in Eq.~\ref{3}. The DLN formalism is simpler since it omits pure photonic degrees of freedoms. Therefore it apparently gives a QED like foundation to the phenomenological model used long time ago by Rytov and Lifshitz for the description of Casimir and van der Walls interactions in term of fluctuating currents~\cite{Lifshitz1956,Ginzburg,Rytov,Callen1951}. In turn, we now  obtain several nonequivalent representations of the physical problem corresponding to the different alternative choices for the Green functions (i.e, retarded, advanced or others). In other words, the DLN method explicitly breaks time symmetry which is a price to pay for its effectiveness and simplicity during calculations. Clearly, something should be added to the DLN formalism  in order to preserve unitarity and time symmetry and thus keeping the symmetric role of $\mathbf{E}_{\textrm{in}}^{(\textrm{eff.})}(\mathbf{x},t)$ and $\mathbf{E}_{\textrm{ret.}}^{(\textrm{eff.})}(\mathbf{x},t)$ needed in any self consistent Hamiltonian approach of electrodynamics and QNP.\\
\indent In order to clarify this issue we must discuss more carefully the role of the bulk medium in the DLN analysis. We will show through this discussion how to remove the ambiguities and limitations of the presently accepted DLN formalism and therefore demonstrate a practical equivalence between the full Huttner-Barnett Hamiltonian description of Sec. II A and an alternative approach generalizing the DLN method originally developed in~\cite{Gruner1995,Gruner1996,Yeung1996}. 
\section{Effective equivalence between the Langevin noise approach and the Huttner-Barnett Hamiltonian description}
\subsection{A more rigorous definition of the effective surrounding medium}
\indent For the present study we first consider a  dielectric medium such that the linear local susceptibility~\cite{A} $2\pi\tilde{\chi}_\omega(\mathbf{x})=\varepsilon_{\omega}(\mathbf{x})-1$ is split into two contributions $\tilde{\chi}^{(1+2)}_\omega(\mathbf{x})=\tilde{\chi}^{(1)}_\omega(\mathbf{x})+\tilde{\chi}^{(2)}_\omega(\mathbf{x})$. In \cite{B} we showed that Eq.~\ref{4} reads
\begin{eqnarray}
\widetilde{\mathbf{E}}_\omega(\mathbf{x})=\widetilde{\mathbf{E}}_{\textrm{in},\omega}^{(\textrm{eff.,1+2})}(\mathbf{x})+\int d^3 \mathbf{x'}\frac{\omega^2}{c^2}\mathbf{G}_\omega^{(\textrm{eff.,1+2})}(\mathbf{x},\mathbf{x'})\nonumber\\ \cdot\widetilde{\mathbf{P}}^{(\textrm{eff.})}_\omega(\mathbf{x'}) \label{4c}
\end{eqnarray} with  the hierarchy
\begin{eqnarray}
\widetilde{\mathbf{E}}_{\textrm{in},\omega}^{(\textrm{eff.,1+2})}(\mathbf{x})=\widetilde{\mathbf{E}}_{\textrm{in},\omega}^{(\textrm{eff.,1})}(\mathbf{x})\nonumber\\+\int d^3 \mathbf{u}\frac{\omega^2}{c^2}\mathbf{G}_{\omega}^{(\textrm{eff.,1})}(\mathbf{x},\mathbf{u}) \cdot 2\pi\tilde{\chi}^{(2)}_\omega(\mathbf{u})\widetilde{\mathbf{E}}_{\textrm{in},\omega}^{(\textrm{eff.,1+2})}(\mathbf{u})\nonumber\\
\mathbf{G}_\omega^{(\textrm{eff.,1+2})}(\mathbf{x},\mathbf{x'})=\mathbf{G}_\omega^{(\textrm{eff.,1})}(\mathbf{x},\mathbf{x'})\nonumber\\ +\int d^3 \mathbf{u}\frac{\omega^2}{c^2}\mathbf{G}_\omega^{(\textrm{eff.,1})}(\mathbf{x},\mathbf{u})\cdot2\pi\tilde{\chi}^{(2)}_\omega(\mathbf{u})\mathbf{G}_\omega^{(\textrm{eff.,1+2})}(\mathbf{u},\mathbf{x'})\nonumber\\ 
\label{7b}
\end{eqnarray}
and 
\begin{eqnarray}
\widetilde{\mathbf{E}}_{\textrm{in},\omega}^{(\textrm{eff.,1})}(\mathbf{x})=\widetilde{\mathbf{E}}_{\textrm{in},\omega}^{(v)}(\mathbf{x})\nonumber\\+\int d^3 \mathbf{u}\frac{\omega^2}{c^2}\mathbf{G}_{\omega}^{(v)}(\mathbf{x},\mathbf{u}) \cdot 2\pi\tilde{\chi}^{(1)}_\omega(\mathbf{u})\widetilde{\mathbf{E}}_{\textrm{in},\omega}^{(\textrm{eff.,1})}(\mathbf{u})\nonumber\\
\mathbf{G}_\omega^{(\textrm{eff.,1})}(\mathbf{x},\mathbf{x'})=\mathbf{G}_\omega^{(v)}(\mathbf{x},\mathbf{x'})\nonumber\\ +\int d^3 \mathbf{u}\frac{\omega^2}{c^2}\mathbf{G}_\omega^{(v)}(\mathbf{x},\mathbf{u})\cdot2\pi\tilde{\chi}^{(1)}_\omega(\mathbf{u})\mathbf{G}_\omega^{(\textrm{eff.,1})}(\mathbf{u},\mathbf{x'})\nonumber\\ 
\label{7c}
\end{eqnarray}
 In defining Eq.~\ref{7c}, which is reminiscent of Eq.~1, we introduced the medium of permittivity  $\varepsilon^{(1)}_{\omega}(\mathbf{x})=2\pi\tilde{\chi}^{(1)}_\omega(\mathbf{x})+1$ as immersed in vacuum while in Eq.~\ref{7b} we constructed an effective medium $1+2$ by adding a susceptibility $2\pi\tilde{\chi}^{(2)}_\omega(\mathbf{x})$ immersed in the background  1 of susceptibility  $2\pi\tilde{\chi}^{(1)}_\omega(\mathbf{x})$.\\
\indent For the present problem we now consider as background medium 1 a quasi-homogeneous susceptibility  in a large volume $V_1$, i.e., such that $2\pi\tilde{\chi}^{(1)}_\omega(\mathbf{x})\simeq 2\pi\tilde{\chi}^{(1)}_\omega$ is spatially independent of the position vector $\mathbf{x}\in V_1$, while  $2\pi\tilde{\chi}^{(1)}_\omega(\mathbf{x})\simeq 0$ for points $\mathbf{x}$ outside $V_1$. With such a choice the field $\widetilde{\mathbf{E}}_{\textrm{in},\omega}^{(\textrm{eff.,1})}(\mathbf{x})\simeq 0$ with an arbitrary large precision for any point $\mathbf{x}\in V_1$ if $V_1\rightarrow+ \infty$ is large enough.
\begin{figure}[]
\centering
\includegraphics[width=0.7\columnwidth]{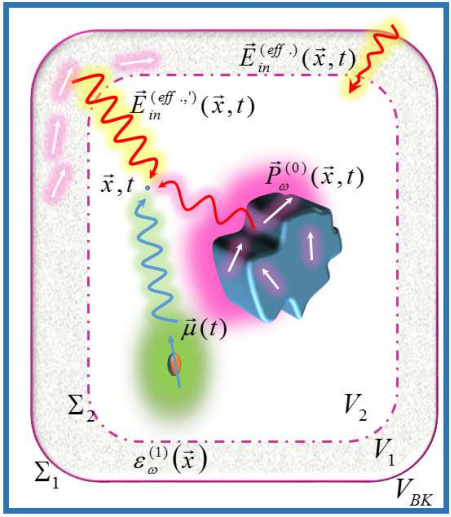}
\caption{Sketch of the amended DLN formulation of the problem shown in Fig.~1.  The system is filled with a weakly dissipative bulk medium with permittivity $\varepsilon^{(1)}_{\omega}(\mathbf{x})$ extending to infinity  (the volume $V_1\rightarrow V_{\textrm{BK}} $) and absorbing any pure and scattered  photon modes $\mathbf{E}_{\textrm{in}}^{(\textrm{eff.})}(\mathbf{x},t)$. The physical system considered in Fig. 1(A) is located near the center of the  large empty region  of volume $V_2\subset V_1$. The dipole distribution located in the region $V_1-V_2$ (i.e., corresponding to the far-field of the physical system in $V_2$) acts as a source of effective photons $\mathbf{E}_{\textrm{in}}^{(\textrm{eff.},')}(\mathbf{x},t)$ having all the properties of the pure photon field  
$\mathbf{E}_{\textrm{in}}^{(\textrm{eff.})}(\mathbf{x},t)$ of Fig.~1(A).  } \label{figure2}
\end{figure}
Physically speaking, this is justified since the incident waves coming from infinity, and characterized by the field $\widetilde{\mathbf{E}}_\omega^{(v)}(\mathbf{x})$, are weakly reflected by the medium (the typical reflection Fresnel coefficient at the boundary $\Sigma_1$ surrounding $V_1$ is $R\sim \frac{\sqrt{\varepsilon^{(1)}_{\omega}}-1}{\sqrt{\varepsilon^{(1)}_{\omega}}+1}\rightarrow 0$ and the transmission coefficient is $T\sim\frac{2\sqrt{\varepsilon^{(1)}_{\omega}}}{\sqrt{\varepsilon^{(1)}_{\omega}}+1}\rightarrow 1$). However, the transmitted waves are  always exponentially damped in the  causal medium 1 due to losses and the resulting field $\widetilde{\mathbf{E}}_{\textrm{in},\omega}^{(\textrm{eff.,1})}(\mathbf{x})\simeq 0$ therefore cancels for points $\mathbf{x}\in V_1$ located sufficiently far apart from the boundary $\Sigma_1=\partial V_1$ surrounding $V_1$. Here we will suppose  that we work exclusively in this regime and we will furthermore  add the hypothesis that  $\varepsilon_{\omega}^{(1)}\rightarrow 1+i0^+$ meaning that the volume $V_1$ has to be very large in order to get $\widetilde{\mathbf{E}}_{\textrm{in},\omega}^{(\textrm{eff.,1})}(\mathbf{x})\simeq 0$.\\ 
\indent In the next step, we insert in the medium 1 a inhomogeneous distribution of  dielectric matter characterized by $2\pi\tilde{\chi}^{(2)}_\omega(\mathbf{x})$ and  we also consider external molecular dipoles  with distribution  $\textbf{P}^{(\textrm{mol.})}(\mathbf{x},t)$. All these systems are supposed to be far away from the boundary $\Sigma_1$ and for definiteness we will consider that all the  points $\textbf{x}$ and systems of interest are located in the volume $V_2<< V_1$. More specifically in order to define the medium 2 we introduce in the volume $V_1$ a large void of volume $V_2$ containing all relevant molecular sources $\textbf{P}^{(\textrm{mol.})}(\mathbf{x},t)$ and the relevant localized dielectric objects of susceptibility $2\pi\tilde{\chi}^{(3)}_\omega(\mathbf{x})$. Furthermore, in this model  all the points $\textbf{x}$ of physical interest and the material systems including the distribution $2\pi\tilde{\chi}^{(3)}_\omega(\mathbf{x})$ and the molecular dipoles $\textbf{P}^{(\textrm{mol.})}(\mathbf{x},t)$ are far apart from the boundary $\Sigma_2=\partial V_2$ surrounding $V_2$ (see Fig.~2). In such a problem the medium 1 is located infinitely far away from the physical systems and can be interpreted  as an absorber modeling the rest of the universe (this is reminiscent of the absorber introduced by Wheeler and Feynman but the strategy used by them was clearly different).  We thus choose as susceptibility $2\pi\tilde{\chi}^{(2)}_\omega(\mathbf{x})$
\begin{eqnarray}
2\pi\tilde{\chi}^{(2)}_\omega(\mathbf{x})=-2\pi\tilde{\chi}^{(1)}_\omega(\mathbf{x})+2\pi\tilde{\chi}^{(3)}_\omega(\mathbf{x}) &\textrm{if $\mathbf{x}\in V_2$,}
\label{8}
\end{eqnarray}  and $2\pi\tilde{\chi}^{(2)}_\omega(\mathbf{x})=0$ otherwise. If we consider $\tilde{\chi}^{(1+2)}_\omega(\mathbf{x})$ we conclude that the term with the minus sign in Eq.~\ref{8} exactly compensates the susceptibility $2\pi\tilde{\chi}^{(1)}_\omega$ for $\mathbf{x}\in V_2$ and therefore at the end the resulting  material system 1+2 located in $V_2$ contains only molecular dipoles $\textbf{P}^{(\textrm{mol.})}(\mathbf{x},t)$ and an inhomogeneous dielectric medium with local permittivity  $\varepsilon^{(3)}_{\omega}(\mathbf{x})=2\pi\tilde{\chi}^{(3)}_\omega(\mathbf{x})+1$.\\ 
\indent Going back to Eq.~\ref{4c} for the total system 1+2 this suggests us  to rewrite:
\begin{eqnarray}
\widetilde{\mathbf{E}}_\omega(\mathbf{x})=\widetilde{\mathbf{E}}_{\textrm{in},\omega}^{(\textrm{eff.,1+2,'})}(\mathbf{x})\nonumber\\+\int_{V_2} d^3 \mathbf{x'}\frac{\omega^2}{c^2}\mathbf{G}_\omega^{(\textrm{eff.,1+2})}(\mathbf{x},\mathbf{x'}) \cdot\widetilde{\mathbf{P}}^{(\textrm{eff.})}_\omega(\mathbf{x'}), \label{9}
\end{eqnarray} with the new effective field 
\begin{eqnarray}
\widetilde{\mathbf{E}}_{\textrm{in},\omega}^{(\textrm{eff.,1+2,'})}(\mathbf{x})=\widetilde{\mathbf{E}}_{\textrm{in},\omega}^{(\textrm{eff.,1+2})}(\mathbf{x})\nonumber\\+\int_{V_1-V_2} d^3 \mathbf{x'}\frac{\omega^2}{c^2}\mathbf{G}_\omega^{(\textrm{eff.,1+2})}(\mathbf{x},\mathbf{x'}) \cdot\widetilde{\mathbf{P}}^{(0)}_\omega(\mathbf{x'}). \label{10}
\end{eqnarray} in which the integration is taken over the complementary volume $V_1-V_2$.\\
\indent Moreover, from its definition in Eq.~\ref{10} $\widetilde{\mathbf{E}}_{\textrm{in},\omega}^{(\textrm{eff.,1+2,'})}(\mathbf{x})$ fulfills homogeneous Maxwell's equations in a dielectric medium with permittivity $\varepsilon^{(3)}_{\omega}(\mathbf{x})$ for any points  $\mathbf{x}\in V_2$. This suggests to interpret this field as an effective photon field. Furthermore, from the two recursive relations in  Eq.~\ref{7b} we can rewrite Eq.~\ref{10} as
\begin{eqnarray}
\widetilde{\mathbf{E}}_{\textrm{in},\omega}^{(\textrm{eff.,1+2,'})}(\mathbf{x})=\widetilde{\mathbf{E}}_{\textrm{in},\omega}^{(\textrm{eff.,1,'})}(\mathbf{x})\nonumber\\+\int d^3 \mathbf{u}\frac{\omega^2}{c^2}\mathbf{G}_{\omega}^{(\textrm{eff.,1+2})}(\mathbf{x},\mathbf{u}) \cdot 2\pi\tilde{\chi}^{(2)}_\omega(\mathbf{u})\widetilde{\mathbf{E}}_{\textrm{in},\omega}^{(\textrm{eff.,1,'})}(\mathbf{u}),\nonumber\\
\label{12}
\end{eqnarray}
with the new field variable
\begin{eqnarray}
\widetilde{\mathbf{E}}_{\textrm{in},\omega}^{(\textrm{eff.,1,'})}(\mathbf{x})=\widetilde{\mathbf{E}}_{\textrm{in},\omega}^{(\textrm{eff.,1})}(\mathbf{x})\nonumber\\+\int_{V_1-V_2} d^3 \mathbf{x'}\frac{\omega^2}{c^2}\mathbf{G}_\omega^{(\textrm{eff.,1})}(\mathbf{x},\mathbf{x'}) \cdot\widetilde{\mathbf{P}}^{(0)}_\omega(\mathbf{x'}). \label{10b}
\end{eqnarray}
Eq.~\ref{12} is formally identical to Eq.~\ref{7b} if we omit the `prime' symbol. This corresponds to the difference of definitions used for $\widetilde{\mathbf{E}}_{\textrm{in},\omega}^{(\textrm{eff.,1,'})}(\mathbf{x})$  and   $\widetilde{\mathbf{E}}_{\textrm{in},\omega}^{(\textrm{eff.,1})}(\mathbf{x})$ respectively. We emphasize that  while $\widetilde{\mathbf{E}}_{\textrm{in},\omega}^{(\textrm{eff.,1})}(\mathbf{x})$ is intrinsically connected to the knowledge of the photon operator $c_{\alpha,j}^{(v)}$, and $c_{\alpha,j}^{(v)\dagger}$ in vacuum the alternative field  $\widetilde{\mathbf{E}}_{\textrm{in},\omega}^{(\textrm{eff.,1,'})}(\mathbf{x})$ additionally introduces an independent contribution from the dipole density $\widetilde{\mathbf{P}}^{(0)}_\omega$  in the volume $V_1-V_2$ so that these fields are not rigorously equivalent.\\
\indent We observe that for the system considered here the condition $\widetilde{\mathbf{E}}_{\textrm{in},\omega}^{(\textrm{eff.,1})}(\mathbf{x})\simeq 0$ $\mathbf{x}\in V_1$ implies (i.e., from Eq.~\ref{7b}) $\widetilde{\mathbf{E}}_{\textrm{in},\omega}^{(\textrm{eff.,1+2})}(\mathbf{x})\simeq 0$ in the same volume $V_1$. Therefore, the field $\widetilde{\mathbf{E}}_{\textrm{in},\omega}^{(\textrm{eff.,1+2,'})}(\mathbf{x})$ is with a very good approximation calculated as
 \begin{eqnarray}
\widetilde{\mathbf{E}}_{\textrm{in},\omega}^{(\textrm{eff.,1+2,'})}(\mathbf{x})\simeq\int_{V_1-V_2} d^3 \mathbf{x'}\frac{\omega^2}{c^2}\mathbf{G}_\omega^{(\textrm{eff.,1+2})}(\mathbf{x},\mathbf{x'}) \nonumber\\ \cdot\widetilde{\mathbf{P}}^{(0)}_\omega(\mathbf{x'}), \label{11}
\end{eqnarray}
 and similarly 
\begin{eqnarray}
\widetilde{\mathbf{E}}_{\textrm{in},\omega}^{(\textrm{eff.,1,'})}(\mathbf{x})\simeq\int_{V_1-V_2} d^3 \mathbf{x'}\frac{\omega^2}{c^2}\mathbf{G}_\omega^{(\textrm{eff.,1})}(\mathbf{x},\mathbf{x'}) \cdot\widetilde{\mathbf{P}}^{(0)}_\omega(\mathbf{x'}),\nonumber\\ \label{10c}
\end{eqnarray} which now depends only on the dipole density $\widetilde{\mathbf{P}}^{(0)}_\omega$  in the volume $V_1-V_2$ and not anymore on the free photon operators.\\
\indent All this discussion was done in order to remove the field  $\widetilde{\mathbf{E}}_{\textrm{in},\omega}^{(\textrm{eff.,1+2})}(\mathbf{x})$ and to consider instead the effective field $\widetilde{\mathbf{E}}_{\textrm{in},\omega}^{(\textrm{eff.,1+2,'})}(\mathbf{x})$. Now if we go back to Eq.~\ref{7b} for the Green tensor in the full medium 1+2 we have 
\begin{eqnarray}
\mathbf{G}_\omega^{(\textrm{eff.,1+2})}(\mathbf{x},\mathbf{x'})=\mathbf{G}_\omega^{(\textrm{eff.,1})}(\mathbf{x},\mathbf{x'})\nonumber\\ +\int_{V_2} d^3 \mathbf{u}\frac{\omega^2}{c^2}\mathbf{G}_\omega^{(\textrm{eff.,1+2})}(\mathbf{x},\mathbf{u})\cdot 2\pi\tilde{\chi}^{(3)}_\omega(\mathbf{u})\mathbf{G}_\omega^{(\textrm{eff.,1})}(\mathbf{u},\mathbf{x'})\nonumber\\ -\int_{V_2} d^3 \mathbf{u}\frac{\omega^2}{c^2}\mathbf{G}_\omega^{(\textrm{eff.,1+2})}(\mathbf{x},\mathbf{u})\cdot 2\pi\tilde{\chi}^{(1)}_\omega(\mathbf{u})\mathbf{G}_\omega^{(\textrm{eff.,1})}(\mathbf{u},\mathbf{x'}).\nonumber\\ \label{5new}
\end{eqnarray}
However, since $\tilde{\chi}^{(1)}_\omega\rightarrow 0^+$ the last term  in Eq.~\ref{5new} is negligible compared to the two other terms. Therefore  for $\mathbf{x}\in V_2$ we get 
\begin{eqnarray}
\mathbf{G}_\omega^{(\textrm{eff.,1+2})}(\mathbf{x},\mathbf{x'})\simeq \mathbf{G}_\omega^{(\textrm{eff.,1})}(\mathbf{x},\mathbf{x'})\nonumber\\ +\int_{V_2} d^3 \mathbf{u}\frac{\omega^2}{c^2}\mathbf{G}_\omega^{(\textrm{eff.,1+2})}(\mathbf{x},\mathbf{u})\cdot 2\pi\tilde{\chi}^{(3)}_\omega(\mathbf{u})\mathbf{G}_\omega^{(\textrm{eff.,1})}(\mathbf{u},\mathbf{x'}).\nonumber\\ \label{13}
\end{eqnarray}
This is exactly the integral definition of the Green tensor $\mathbf{G}_\omega^{(\textrm{eff.,3})}(\mathbf{x},\mathbf{x'})$ obtained in presence of the dielectric medium with permittivity $\tilde{\chi}^{(3)}_\omega$ without the surrounding   medium 1 with susceptibility  $2\pi\tilde{\chi}^{(1)}_\omega$:
\begin{eqnarray}
\mathbf{G}_\omega^{(\textrm{eff.,3})}(\mathbf{x},\mathbf{x'})= \mathbf{G}_\omega^{(v)}(\mathbf{x},\mathbf{x'})\nonumber\\ +\int_{V_2} d^3 \mathbf{u}\frac{\omega^2}{c^2}\mathbf{G}_\omega^{(v)}(\mathbf{x},\mathbf{u})\cdot 2\pi\tilde{\chi}^{(3)}_\omega(\mathbf{u})\mathbf{G}_\omega^{(\textrm{eff.,3})}(\mathbf{u},\mathbf{x'}).\nonumber\\ \label{14}
\end{eqnarray}
Therefore, for $\mathbf{x}\in V_2$, we can rewrite Eq.~\ref{10} as
\begin{eqnarray}
\widetilde{\mathbf{E}}_{\textrm{in},\omega}^{(\textrm{eff.,1+2,'})}(\mathbf{x})\simeq\widetilde{\mathbf{E}}_{\textrm{in},\omega}^{(\textrm{eff.,1,'})}(\mathbf{x})\nonumber\\+\int_{V_2} d^3 \mathbf{u}\frac{\omega^2}{c^2}\mathbf{G}_{\omega}^{(\textrm{eff.,3})}(\mathbf{x},\mathbf{u}) \cdot 2\pi\tilde{\chi}^{(3)}_\omega(\mathbf{u})\widetilde{\mathbf{E}}_{\textrm{in},\omega}^{(\textrm{eff.,1,'})}(\mathbf{u}),\nonumber\\
\label{12new}
\end{eqnarray}
and Eq.~\ref{9} as 
\begin{eqnarray}
\widetilde{\mathbf{E}}_\omega(\mathbf{x})\simeq \widetilde{\mathbf{E}}_{\textrm{in},\omega}^{(\textrm{eff.,1+2,'})}(\mathbf{x})\nonumber\\+\int_{V_2} d^3 \mathbf{x'}\frac{\omega^2}{c^2}\mathbf{G}_\omega^{(\textrm{eff.,3})}(\mathbf{x},\mathbf{x'}) \cdot\widetilde{\mathbf{P}}^{(\textrm{eff.})}_\omega(\mathbf{x'}). \label{9new}
\end{eqnarray}
The two last equations  are very similar to the results we would obtain for the description of the total field in presence of the dielectric medium 3 alone, i.e. without the surrounding medium 1. For this different problem we indeed have
   \begin{eqnarray}
\widetilde{\mathbf{E}}_{\textrm{in},\omega}^{(\textrm{eff.,3})}(\mathbf{x})=\widetilde{\mathbf{E}}_{\textrm{in},\omega}^{(v)}(\mathbf{x})\nonumber\\+\int_{V_2} d^3 \mathbf{u}\frac{\omega^2}{c^2}\mathbf{G}_{\omega}^{(\textrm{eff.,3})}(\mathbf{x},\mathbf{u}) \cdot 2\pi\tilde{\chi}^{(3)}_\omega(\mathbf{u})\widetilde{\mathbf{E}}_{\textrm{in},\omega}^{(\textrm{eff.,3})}(\mathbf{u}),\nonumber\\
\label{12c}
\end{eqnarray}
and  
\begin{eqnarray}
\widetilde{\mathbf{E}}_\omega(\mathbf{x})= \widetilde{\mathbf{E}}_{\textrm{in},\omega}^{(\textrm{eff.,3})}(\mathbf{x})\nonumber\\+\int_{V_2} d^3 \mathbf{x'}\frac{\omega^2}{c^2}\mathbf{G}_\omega^{(\textrm{eff.,3})}(\mathbf{x},\mathbf{x'}) \cdot\widetilde{\mathbf{P}}^{(\textrm{eff.})}_\omega(\mathbf{x'}). \label{9c}
\end{eqnarray}  
 Formally, the equivalence would be complete if we could write $\widetilde{\mathbf{E}}_{\textrm{in},\omega}^{(\textrm{eff.,1,'})}(\mathbf{x})=\widetilde{\mathbf{E}}_{\textrm{in},\omega}^{(v)}(\mathbf{x})$. 
This is of course not rigorously possible since $\widetilde{\mathbf{E}}_{\textrm{in},\omega}^{(\textrm{eff.,1,'})}(\mathbf{x})$ is a solution of homogenous Maxwell's  equation in the bulk medium  1 while 
$\widetilde{\mathbf{E}}_{\textrm{in},\omega}^{(v)}(\mathbf{x})$ is a solution of homogeneous Maxwell's equations in vacuum. Also, from the QED or QNP point of view the operators are not acting on the  same Hilbert spaces since one field acts on the pure material oscillator states while the other acts on the pure photon states. Still, since  $\tilde{\chi}^{(1)}_\omega\rightarrow 0^+$ we must show that these problems are not fundamental for a practical perspective. In order to do that we have to consider more in details the dynamics and the commutation  relations associated with the  electric field operator $\widetilde{\mathbf{E}}_{\textrm{in},\omega}^{(\textrm{eff.,1,'})}(\mathbf{x})$ to see that we can indeed consider this field as describing a kind of effective photon field.  \\
\subsection{Effective photon field}
\indent We remind that  in \cite{B} we studied the problem of the homogeneous bulk medium in details using the Laplace transform method for solving Maxwell's equations in the Heisenberg representation.  We showed that in the limit where this medium 1 with susceptibility $2\pi\tilde{\chi}^{(1)}_\omega$ is infinitely extended (i.e., filling the full Born von Karman volume $V_{\textrm{BK}}$) and in absence of molecular dipoles, i.e, $\textbf{P}^{(\textrm{mol.})}(\mathbf{x},t)=0$, we can split the retarded field  
\begin{eqnarray}\widetilde{\mathbf{E}}_{\textrm{ret.},\omega}^{(\textrm{eff.,1})}(\mathbf{x})=\int_{V_{\textrm{BK}}} d^3 \mathbf{x'}\frac{\omega^2}{c^2}\mathbf{G}_\omega^{(\textrm{eff.,1})}(\mathbf{x},\mathbf{x'})\cdot\widetilde{\mathbf{P}}^{(0)}_\omega(\mathbf{x'})\label{retarded}\end{eqnarray}
 into a purely transverse (i.e., solenoidal) field $\widetilde{\mathbf{E}}_{\textrm{ret.},\bot,\omega}^{(\textrm{eff.,1})}(\mathbf{x})$ and into a purely longitudinal (i.e., irrotational) field $\widetilde{\mathbf{E}}_{\textrm{ret.},||,\omega}^{(\textrm{eff.,1})}(\mathbf{x})$ such that for $\omega>0$: 
\begin{eqnarray}
\widetilde{\mathbf{E}}_{\textrm{ret.},\bot,\omega}^{(\textrm{eff.,1})}(\mathbf{x})=\sum_{\alpha,j}\frac{\omega^2\mathbf{E}_{\alpha,j}^{(v)}(\mathbf{x}) }{\omega_\alpha^2-\omega^2\varepsilon_\omega^{(1)}}\sqrt{\frac{2\hbar\varepsilon_\omega^{'',(1)}}{\pi}}\frac{f^{(0)}_{\omega,\alpha,j}(t_0)e^{i\omega_\alpha t_0}}{\sqrt{\hbar \omega_\alpha}},\nonumber\\ 
\widetilde{\mathbf{E}}_{\textrm{ret.},||,\omega}^{(\textrm{eff.,1})}(\mathbf{x})=-\sum_{\alpha}\frac{e^{i\mathbf{k}_\alpha\cdot\mathbf{x}}\boldsymbol{\hat{k}}_{\alpha}}{\sqrt{V_{\textrm{BK}}}\varepsilon_\omega^{(1)}}\sqrt{\frac{\hbar\varepsilon_\omega^{'',(1)}}{\pi}}f^{(0)}_{\omega,\alpha,||}(t_0)e^{i\omega_\alpha t_0},\nonumber\\
\label{reglop} 
\end{eqnarray} and for $\omega<0$ we have $\widetilde{\mathbf{E}}_{\textrm{ret.},\omega}^{(\textrm{eff.,1})}(\mathbf{x})=\widetilde{\mathbf{E}}_{\textrm{ret.},-\omega}^{\dagger(\textrm{eff.,1})}(\mathbf{x})$. In Eq.~\ref{reglop} we introduced \cite{B} the lowering operators $f^{(0)}_{\omega,\alpha,j}(t)=\int_{V_{\textrm{BK}}} d^3\mathbf{x'}\sqrt{\frac{2}{\hbar \omega_\alpha}}\mathbf{E}_{\alpha,j}^{\ast,(v)}(\mathbf{x})\cdot\mathbf{f}^{(0)}_\omega(\mathbf{x},t)$ and $f^{(0)}_{\omega,\alpha,||}(t)=\int_{V_{\textrm{BK}}} d^3\mathbf{x'}\frac{e^{-i\mathbf{k}_\alpha\cdot\mathbf{x}}}{\sqrt{V_{\textrm{BK}}}}\boldsymbol{\hat{k}}_{\alpha}\cdot\mathbf{f}^{(0)}_\omega(\mathbf{x},t)$ satisfying the commutation rules $[f^{(0)}_{\omega,\alpha,j}(t),f^{\dagger(0)}_{\omega',\beta,k}(t)]=\delta_{\alpha,\beta}\delta_{j,k}\delta(\omega-\omega')$ and $[f^{(0)}_{\omega,\alpha,||}(t),f^{\dagger(0)}_{\omega',\beta,||}(t)]=\delta_{\alpha,\beta}\delta_{j,k}\delta(\omega-\omega')$ (the other commutators vanish).\\
\indent What is important in Eq.~\ref{reglop} is the presence of polariton resonances in the frequency domain canceling the denominators of the transverse and longitudinal fields. These resonances occur for frequencies solutions of  $\omega_\alpha^2-\omega^2\varepsilon_\omega^{(1)}=0$ (transverse modes) and $\varepsilon_\omega^{(1)}=0$ (longitudinal modes). Furthermore, since the medium is causal and lossy   the solutions $\Omega$ are generally located in the lower part of the complex plane (i.e., with $\Omega''<0$). In \cite{B} we showed that for a weakly dissipative medium such as a Drude-Lorentz metal it makes sense to define new effective photon annihilation and creation operators labeled by such polaritons modes. For the present purpose we consider the simple  Drude-Lorentz permittivity 
\begin{eqnarray}
\varepsilon_\omega^{(1)}=1+\frac{\omega_P^2}{\omega_0^2-(\omega+i\gamma)^2},
\end{eqnarray} where $\omega_P$,  $\omega_0$, $\gamma$ are real and positive constants. In the limit $\gamma\rightarrow 0^+$ this leads to the Hopfield-Fano polariton model~\cite{Fano1956,Hopfield1958} and we get a longitudinal mode without dispersion $\Omega_{||}(\omega_\alpha)=\omega_L-i\gamma$ ($\omega_L=\sqrt{\omega_P^2+\omega_0^2}$) and two dispersive transverse polariton branches which in the lossless limit are given by  $\Omega_{\bot,\pm}(\omega_\alpha)=\frac{\sqrt{[\omega_\alpha^2+\omega_L^2\pm\sqrt{((\omega_\alpha^2+\omega_L^2)^2-4\omega_\alpha^2\omega_0^2)}]}}{\sqrt{2}}$.
In \cite{B} we showed that the effective photon annihilation operators associated with the transverse electric field are defined by 
\begin{eqnarray}
c_{\alpha,j,\pm}(t)=\frac{1}{N_{\alpha,\pm}}\int_{\delta \Omega_{\alpha,m}} d\omega\frac{\omega^2}{\omega_\alpha^2-\omega^2\varepsilon_\omega^{(1)}}\sqrt{\frac{\hbar\varepsilon_\omega^{''(1)}}{\pi}}f^{(0)}_{\omega,\alpha,j}(t).\nonumber\\ 
\label{modepolar}\end{eqnarray} where $\delta \Omega_{\alpha,m}$ is a frequency window centered on the polariton pulsation $\textrm{Re}[\Omega_{\bot,\pm}(\omega_\alpha)]$ and where $N_{\alpha,\pm}$ is a normalization  constant given by $\sqrt{[\frac{\hbar\Omega_{\bot,\pm}(\omega_\alpha)}{2}\frac{d\Omega{\bot,\pm}(\omega_\alpha)^2}{d\omega_\alpha^2}]}$. These operators satisfy the standard bosonic commutation relations such as $[c_{\alpha,j,\beta}(t),c_{\alpha',j',\beta'}^\dagger(t)]=\delta_{\alpha',\alpha}\delta_{j',j}\delta_{\beta',\beta}$ (with $\beta,\beta'=\pm$) ensuring the interpretation in term of annihilation/creation operators.\\
\indent  Now, in the system we consider we will impose $\omega_0, \omega_P \rightarrow 0 $ so that the lower polariton branch with horizontal asymptote at $\lim_{\omega_\alpha\rightarrow +\infty}[\Omega{\bot,-}(\omega_\alpha)]\simeq \omega_L\rightarrow 0$ will not play any role for frequency $\omega\gg\omega_L$. In this regime the upper polariton branch has a dispersion approaching the asymptote $\Omega_{\bot,+}(\omega_\alpha)\simeq\omega_\alpha $. Therefore, for a large spectral band of frequencies $\omega\gg\omega_L$ corresponding to the physical dielectric excitations (associated with the operators $f^{(0)}_{\omega,\alpha,j}(t_0)$, $f^{\dagger(0)}_{\omega,\alpha,j}(t_0)$) we will find a quasi resonant value $\omega\simeq\omega_\alpha$ where the integrand in Eq.~\ref{modepolar} will be very high. In this regime the bulk medium is quasi transparent and the mode operators have an harmonic evolution $c_{\alpha,j,\pm}(t)\simeq c_{\alpha,j,\pm}(t_0)e^{-i\omega_\alpha (t-t_0)}$. Any realistic  material excitations associated with a pulse of finite width $\Delta\omega$ centered on a frequency $\omega\gg\omega_L$ will thus be described by this dynamics associated with effective photons and the transverse electric field operator. Furthermore, the longitudinal field will not play any role in the formalism since $\Omega_{||}(\omega_\alpha)\simeq \omega_L\rightarrow 0$.\\
\indent We can thus write with a very good approximation $\mathbf{E}_{\textrm{ret.}}^{(\textrm{eff.,1})}(\mathbf{x},t)\simeq \mathbf{E}_{\textrm{ret.},\bot}^{(\textrm{eff.,1})}(\mathbf{x},t)$ with
 \begin{eqnarray}
\mathbf{E}_{\textrm{ret.},\bot}^{(\textrm{eff.,1})}(\mathbf{x})\simeq \sum_{\alpha,j}\mathbf{E}_{\alpha,j}^{(v)}(\mathbf{x})c_{\alpha,j,+}(t_0)e^{-i\omega_\alpha (t-t_0)}+hcc.\nonumber\\
\label{30} 
\end{eqnarray} and  equivalently  
\begin{eqnarray}
\widetilde{\mathbf{E}}_{\textrm{ret.},\bot,\omega}^{(\textrm{eff.,1})}(\mathbf{x})\simeq\sum_{\alpha,j}[\mathbf{E}_{\alpha,j}^{(v)}(\mathbf{x})c_{\alpha,j,+}(t_0)e^{i\omega_\alpha t_0}\delta(\omega-\omega_\alpha)\nonumber\\ + \mathbf{E}_{\alpha,j}^{\ast (v)}(\mathbf{x})c_{\alpha,j,+}(t_0)e^{-i\omega_\alpha t_0}\delta(\omega+\omega_\alpha)].
\label{31} 
\end{eqnarray} 
Eq.~\ref{31} and thus Eq.~\ref{retarded} are  clearly reminiscent of Eq.~\ref{2} for the pure photon field in vacuum.\\
\indent We now go back to Eq.~\ref{10c} for $\widetilde{\mathbf{E}}_{\textrm{in},\omega}^{(\textrm{eff.,1,'})}(\mathbf{x})$  and realize that in the limit $V_1\rightarrow V_{\textrm{BK}}$ and $V_2/V_1\rightarrow 0$ the integral in Eq.~\ref{10c} becomes equivalent to the one in Eq.~\ref{retarded}. Therefore we get  $\widetilde{\mathbf{E}}_{\textrm{in},\omega}^{(\textrm{eff.,1,'})}(\mathbf{x})=\widetilde{\mathbf{E}}_{\textrm{ret.},\omega}^{(\textrm{eff.,1})}(\mathbf{x})$ which from Eq.~\ref{31} allows us to write
\begin{eqnarray}
\widetilde{\mathbf{E}}_{\textrm{in},\omega}^{(\textrm{eff.,1,'})}(\mathbf{x})\simeq\sum_{\alpha,j}[\mathbf{E}_{\alpha,j}^{(v)}(\mathbf{x})c_{\alpha,j,+}(t_0)e^{i\omega_\alpha t_0}\delta(\omega-\omega_\alpha)\nonumber\\ + \mathbf{E}_{\alpha,j}^{\ast (v)}(\mathbf{x})c_{\alpha,j,+}(t_0)e^{-i\omega_\alpha t_0}\delta(\omega+\omega_\alpha)].
\label{31b} 
 \end{eqnarray} Finally, from this result and after comparing Eq.~\ref{12new} and \ref{12c} we deduce that the retarded field $\widetilde{\mathbf{E}}_{\textrm{in},\omega}^{(\textrm{eff.,1+2,'})}(\mathbf{x})$ is formally equivalent to the scattered photon field $\widetilde{\mathbf{E}}_{\textrm{in},\omega}^{(\textrm{eff.,3})}(\mathbf{x})$ but now with effective photon operators $c_{\alpha,j,+}(t_0)$, $c_{\alpha,j,+}^\dagger(t_0)$ replacing the free space photon operators $c_{\alpha,j}^{(v)}(t_0)$, $c_{\alpha,j}^{\dagger(v)}(t_0)$. The two fields do not act on the same Hilbert space but by choosing the initial state conveniently we can map any physical problem from one  model to the other. Therefore, we showed that the description using fluctuating dipolar sources $\widetilde{\mathbf{P}}^{(0)}_\omega(\mathbf{x'})$ located in the surrounding medium  (i.e. in the volume $V_1-V_2$) are for all practical needs equivalent to a model involving an effective photon field. In that sense we can say that we generalized and completed the standard DLN formalism  by including new dipolar sources $\widetilde{\mathbf{P}}^{(0)}_\omega(\mathbf{x})$ located far away from the region of interest (i.e., in the region  $V_1-V_2$) which formally speaking are equivalent to the pure photon field that the usual DLN approach removed.  In other words   we showed that the situation sketched in Fig.~2 (which generalizes the one shown in Fig.~1(B)) is equivalent to the situation represented in Fig.~1(A): This is the central finding of this article.  \\   
\section{Discussions and applications}
\indent We shall now summarize the results obtained insofar. We started by modeling an effective medium 1+2 including molecular dipoles and the dielectric medium of susceptibility $2\pi\tilde{\chi}^{(3)}_\omega(\mathbf{x})$  well localized in a large void of volume $V_2$.  This void is surrounded by a medium 1 of quasi homogeneous susceptibility $2\pi\tilde{\chi}^{(1)}_\omega\rightarrow 0^+$ in a  volume  $V_1-V_2$  with  $V_1$ is much larger than $V_2$ and includes entirely $V_2$. We showed (see Eq.~\ref{9}) that the electric field acting at any point $\mathbf{x}$ near the center of $V_2$ can be separated into a retarded contribution of the dipole distribution $\widetilde{\mathbf{P}}^{(\textrm{eff.})}_\omega(\mathbf{x'})$  in $V_2$ and into a retarded contribution $\widetilde{\mathbf{E}}_{\textrm{in},\omega}^{(\textrm{eff.,1+2,'})}(\mathbf{x})$ associated with the fluctuating dipole $\widetilde{\mathbf{P}}^{(0)}_\omega(\mathbf{x'})$  contained in $V_1-V_2$ (see Eq.~\ref{11}). We showed that this last contribution, which  for $\mathbf{x}\in V_2$ satisfies the homogeneous Maxwell equation in presence of the dielectric medium of permittivity $\varepsilon_\omega^{(3)}(\mathbf{x})$, is formally identical to the  effective photon field $\widetilde{\mathbf{E}}_{\textrm{in},\omega}^{(\textrm{eff.,3})}(\mathbf{x})$ solution of a different physical problem (see Eqs.~\ref{12c},\ref{9c}) in which the same dielectric medium  of permittivity $\varepsilon_\omega^{(3)}(\mathbf{x})$  and the same molecular distribution $\textbf{P}^{(\textrm{mol.})}(\mathbf{x},t)$  as considered previously  are not anymore surrounded  by a large weakly absorbing medium of  susceptibility $2\pi\tilde{\chi}^{(1)}_\omega$ but instead by vacuum. In this new problem free photon states are allowed to propagate and to excite points $\textbf{x}$ near the medium of permittivity $\varepsilon_\omega^{(3)}(\mathbf{x})$ and the field  $\widetilde{\mathbf{E}}_{\textrm{in},\omega}^{(\textrm{eff.,3})}(\mathbf{x})$ corresponds to this scattered component.\\
\indent What is key in this demonstration is that we can, i.e., with as large an accuracy as needed, eliminate any free space photon state and replace it by an equivalent radiated field  originating from dipolar sources $\widetilde{\mathbf{P}}^{(0)}_\omega(\mathbf{x'})$ located very far away from the region of interest. Therefore, we get here a formalism which is able to generalize the standard DLN procedure by replacing free space photons (scattered by the environment) by radiative sources located in the far-field. Now, in many calculations it is much simpler to use this alternative description without real free photon field but using instead this concept of effective free photon generated by fluctuating sources. The reason is that this effective photon field is from Eq.~\ref{11} calculated using the Green tensor $\mathbf{G}_\omega^{(\textrm{eff.,1+2})}(\mathbf{x},\mathbf{x'})$ and we can show that for practical calculations (i.e., for points $\mathbf{x}$, $\mathbf{x'}$ very far from the boundaries $\Sigma_2$, $\Sigma_1$) the results are equivalent to those obtained using the standard DLN method neglecting the surrounding environment of susceptibility $2\pi\tilde{\chi}^{(1)}_\omega$. In order to appreciate this fact further we will now consider few examples of calculations involving correlators and fluctuations for QNP.\\
\subsection{The fundamental commutation relations for QNP} 
\indent In their fundamental articles introducing the DLN approach Gruner and Welsch~\cite{Gruner1995,Gruner1996,Yeung1996} explicitly calculated the canonical commutators involving the electric or magnetic field operators defined at two spatial positions $\textbf{a}$ and $\textbf{b}$ and two different times $t_a$ and $t_b$. These quantities are central for calculating quantum observable  associated with field fluctuations and correlations \cite{Fermani2006}. Here, we will consider specifically the case of the commutator  $[\breve{\textbf{E}}^{(+)}(\textbf{a},t_a),\breve{\textbf{E}}^{(-}(\textbf{b},t_b)]$ involving the positive and negative frequency parts of the electric field operator, which plays a central role in QED.\\            
\indent In order to be clear we should define  precisely what we mean here by positive and negative frequency parts. Following Glauber~\cite{Glauber} we define the positive and respectively negative frequency part of any time dependent operator $F(t)$  as an Hilbert transform \begin{eqnarray}
\breve{F}^{(\pm)}(t)=\pm\int_{-\infty}^{+\infty}\frac{d\tau}{2\pi i} \frac{F(t-\tau)}{\tau \mp i 0^+},\label{pos1}
\end{eqnarray}
 which leads to the standard explicit forms
\begin{eqnarray}
\breve{F}^{(+)}(t)=\int_{0}^{+\infty}d\omega\tilde{F}_\omega e^{-i\omega t},\nonumber\\
\breve{F}^{(-)}(t)=\int_{-\infty}^{0}d\omega\tilde{F}_\omega e^{+i\omega t}\label{pos2}
\end{eqnarray} ensuring $F(t)=\breve{F}^{(+)}(t)+\breve{F}^{(-)}(t)$ [in particular if the operator is Hermitian $F(t)=F(t)^\dagger$ we have $(\breve{F}^{(+)}(t))^\dagger=\breve{F}^{(-)}(t)$). We emphasize that the present definition of positive and negative frequency operators do not  exactly corresponds to the canonical separation  into annihilation and creation operators. In \cite{B,C} we introduced the operator~\cite{Marx}  $\mathcal{L}_t^{(\pm)}=\frac{1}{2}[1\pm\frac{i\partial_t}{c\sqrt{-\boldsymbol{\nabla}^2}}]$ which applied on the displacement field leads to a clean separation of annihilation and destruction operators contributions $\mathbf{D}^{(\pm)}(\mathbf{x},t)=\mathcal{L}_t^{(\pm)}[\mathbf{D}(\mathbf{x},t)]$. The two definitions are actually equivalent in vacuum and they lead in general to similar results in the far-field (see Appendix C and D in \cite{C}).\\
\indent Now, we consider the application of the definition given in Eq.~\ref{pos2} to the electric field operator $\mathbf{E}(\mathbf{x},t)$ in Eq.~\ref{3} and more specifically to the case where $\textbf{P}^{(\textrm{mol.})}(\mathbf{x},t)=0$ so that  $\textbf{P}^{(\textrm{eff.})}(\mathbf{x},t)=\textbf{P}^{(0)}(\mathbf{x},t)$. The commutator $[\breve{\mathbf{E}}^{(+)}(\mathbf{a},t_a),\breve{\mathbf{E}}^{(-)}(\mathbf{b},t_b)]$ is thus given by 
\begin{eqnarray}
[\breve{\mathbf{E}}^{(+)}(\mathbf{a},t_a),\breve{\mathbf{E}}^{(-)}(\mathbf{b},t_b)]\nonumber\\=\int_{0}^{+\infty}\int_{0}^{+\infty}d\omega'd\omega e^{-i\omega' t_a}e^{+i\omega t_b}[\widetilde{\mathbf{E}}_{\omega'}(\mathbf{a}),\widetilde{\mathbf{E}}^{\dagger}_{\omega}(\mathbf{b})].\label{36}
\end{eqnarray}
Furthermore, since the pure photonic degrees of freedom characterized by the operators  $c_{\alpha,j}^{(v)}(t_0)$, $c_{\alpha,j}^{\dagger(v)}(t_0)$ commute with the pure material oscillator degrees of freedom characterized by $\mathbf{f}^{(0)}_{\omega}(\mathbf{x},t_0)$, $\mathbf{f}^{\dagger(0)}_{\omega}(\mathbf{x},t_0)$ \cite{A,B} we can express the electric field commutator of Eq.~\ref{36} as the sum of a contribution $[\breve{\mathbf{E}}_{\textrm{in}}^{(+)(\textrm{eff.})}(\mathbf{a},t_a),\breve{\mathbf{E}}_{\textrm{in}}^{(-)(\textrm{eff.})}(\mathbf{b},t_b)]$ for the photonic field and a contribution $[\breve{\mathbf{E}}_{\textrm{ret.}}^{(+)(\textrm{eff.})}(\mathbf{a},t_a),\breve{\mathbf{E}}_{\textrm{ret.}}^{(-)(\textrm{eff.})}(\mathbf{b},t_b)]$ for the material field.\\
\indent We consider first the pure photonic correlator $[\breve{\mathbf{E}}_{\textrm{in}}^{(+)(\textrm{eff.})}(\mathbf{a},t_a),\breve{\mathbf{E}}_{\textrm{in}}^{(-)(\textrm{eff.})}(\mathbf{b},t_b)]$,  which  from Eq.~\ref{6} reads: 
\begin{eqnarray}
[\breve{\mathbf{E}}_{\textrm{in}}^{(+)(\textrm{eff.})}(\mathbf{a},t_a),\breve{\mathbf{E}}_{\textrm{in}}^{(-)(\textrm{eff.})}(\mathbf{b},t_b)]\nonumber\\=\sum_{\alpha,j}\mathbf{E}_{\alpha,j}^{(\textrm{eff.})}(\mathbf{a})\otimes\mathbf{E}_{\alpha,j}^{\ast(\textrm{eff.})}(\mathbf{b})e^{-i\omega_\alpha(t_a-t_b)}\label{37} 
\end{eqnarray}
In the limit case of the pure vacuum the only contribution  is $[\breve{\mathbf{E}}_{\textrm{in}}^{(+)(v)}(\mathbf{a},t_a),\breve{\mathbf{E}}_{\textrm{in}}^{(-)(v)}(\mathbf{b},_b)]$, which from Eq.~\ref{2} 
 reads:
\begin{eqnarray}
[\breve{\mathbf{E}}_{\textrm{in}}^{(+)(v)}(\mathbf{a},t_a),\breve{\mathbf{E}}_{\textrm{in}}^{(-)(v)}(\mathbf{b},t_b)]\nonumber\\=\sum_{\alpha,j}\mathbf{E}_{\alpha,j}^{(v)}(\mathbf{a})\otimes\mathbf{E}_{\alpha,j}^{\ast(v)}(\mathbf{b})e^{-i\omega_\alpha(t_a-t_b)}.\label{38} 
\end{eqnarray}
A direct calculation shown in Appendix D of \cite{C} demonstrates that this correlator is also expressed as 
\begin{eqnarray}
[\breve{\mathbf{E}}_{\textrm{in}}^{(+)(v)}(\mathbf{a},t_a),\breve{\mathbf{E}}_{\textrm{in}}^{(-)(v)}(\mathbf{b},t_b)]\nonumber\\
=\int_{0}^{+\infty}d\omega\frac{\hbar\omega}{\pi}\frac{\omega^2}{c^2}\textrm{Imag}[\mathbf{G}_\omega^{(v)}(\mathbf{a},\mathbf{b})]e^{-i\omega(t_a-t_b)}\nonumber\\
=-i\hbar\boldsymbol{\Delta}_{\textrm{ret.}}^{(+)(v)}(|t_a-t_b|,\mathbf{a},\mathbf{b}),
\label{39} 
\end{eqnarray} where $\boldsymbol{\Delta}_{\textrm{ret.}}^{(+)(v)}(\tau,\mathbf{a},\mathbf{b})=\mathcal{L}_\tau^{(\pm)}[\boldsymbol{\Delta}_{\textrm{ret.}}^{(v)}(\tau,\mathbf{a},\mathbf{b})]$, i.e,
\begin{eqnarray}
\boldsymbol{\Delta}_{\textrm{ret.}}^{(+)(v)}(\tau,\mathbf{a},\mathbf{b})\nonumber\\
=\frac{i}{\hbar}\sum_{\alpha,j}\mathbf{E}_{\alpha,j}^{(v)}(\mathbf{a})\otimes\mathbf{E}_{\alpha,j}^{\ast(v)}(\mathbf{b})e^{-i\omega_\alpha\tau}\Theta(\tau)\nonumber\\
=i\int_{0}^{+\infty}d\omega\frac{\hbar\omega}{\pi}\frac{\omega^2}{c^2}\textrm{Imag}[\mathbf{G}_\omega^{(v)}(\mathbf{a},\mathbf{b})]e^{-i\omega \tau}\Theta(\tau).
\label{39b} 
\end{eqnarray}
The integral formula in Eq.~\ref{39} is particularly interesting since  as we will see below it is very similar to the expression obtained  for the material term $[\breve{\mathbf{E}}_{\textrm{ret.}}^{(+)(\textrm{eff.})}(\mathbf{a},t_a),\breve{\mathbf{E}}_{\textrm{ret.}}^{(-)(\textrm{eff.})}(\mathbf{b},t_b)]$ in the context of the DLN formalism.\\
\indent More precisely, in order to calculate the commutator $[\breve{\mathbf{E}}_{\textrm{ret.}}^{(+)(\textrm{eff.})}(\mathbf{a},t_a),\breve{\mathbf{E}}_{\textrm{ret.}}^{(-)(\textrm{eff.})}(\mathbf{b},t_b)]$ we insert into Eq.~\ref{37} the definition for $\widetilde{\mathbf{E}}_{\textrm{ret.}\omega}^{(\textrm{eff.})}(\mathbf{x})$ given by Eq.~\ref{4} and use the definition 
\begin{eqnarray}
\widetilde{\mathbf{P}}^{(0)}_\omega(\mathbf{x})=\sqrt{\frac{\hbar\varepsilon^{''}_{\omega}(\mathbf{x})}{\pi}}\mathbf{f}^{(0)}_{\omega}(\mathbf{x},t_0)e^{i\omega t_0}\theta(\omega)\nonumber\\
+\sqrt{\frac{\hbar\varepsilon^{''}_{-\omega}(\mathbf{x})}{\pi}}\mathbf{f}^{\dagger(0)}_{-\omega}(\mathbf{x},t_0)e^{i\omega t_0}\theta(-\omega),\label{dipole}
\end{eqnarray} which together with the canonical commutations for the $\mathbf{f}^{(0)}_{\omega}$, $\mathbf{f}^{\dagger(0)}_{\omega}$ operators leads to 
\begin{eqnarray}
[\breve{\mathbf{E}}_{\textrm{ret.}}^{(+)(\textrm{eff.})}(\mathbf{a},t_a),\breve{\mathbf{E}}_{\textrm{ret.}}^{(-)(\textrm{eff.})}(\mathbf{b},t_b)]\nonumber\\
=\frac{\hbar}{\pi}\int_0^{+\infty}d\omega\frac{\omega^2}{c^2}\textbf{N}_\omega^{(\textrm{eff.})}(\mathbf{a},\mathbf{b})e^{-i\omega(t_a-t_b)},
\label{39c} 
\end{eqnarray}  
 with 
\begin{eqnarray}
\textbf{N}_\omega^{(\textrm{eff.})}(\mathbf{a},\mathbf{b})=\int d^3\mathbf{x}\frac{\omega^2}{c^2}\varepsilon''_{\omega}(\mathbf{x})\mathbf{G}_\omega^{(\textrm{eff.})}(\mathbf{a},\mathbf{x})\nonumber\\ \cdot\mathbf{G}_\omega^{\ast,(\textrm{eff.})}(\mathbf{x},\mathbf{b}).\label{matt}
\end{eqnarray}
The integral term $\textbf{N}_\omega^{(\textrm{eff.})}(\mathbf{a},\mathbf{b})$ has been evaluated by authors of the DLN formalism~\cite{Gruner1995,Gruner1996,Yeung1996,Scheelreview2008} by using some Green integral identities together with the assumption that the permittivity $\varepsilon_{\omega}^{(\textrm{bulk})}$ at spatial infinity corresponds to an absorbing media (we remind that this is a key issue in DLN formalism). For the present purpose this assumption is not justified, and  we will for generality relax this condition in order to allow the configuration $\varepsilon_{\omega}^{(\textrm{bulk})}=1$.
The details of the calculations based on the dyadic-dyadic Green theorem are given in Appendix A and we get after some manipulations 
\begin{eqnarray}
\textrm{Imag}[\mathbf{G}_\omega^{(\textrm{eff.})}(\mathbf{a},\mathbf{b})]-\oint_{\Sigma_\infty}dS\mathbf{F}_\omega^{(\textrm{eff.})}(\mathbf{x},\mathbf{a},\mathbf{b})\nonumber\\
=\int_{V_\infty}d^3\mathbf{x}\frac{\omega^2}{c^2}\varepsilon''_{\omega}(\mathbf{x})\mathbf{G}_\omega^{(\textrm{eff.})}(\mathbf{a},\mathbf{x})\cdot\mathbf{G}_\omega^{\ast,(\textrm{eff.})}(\mathbf{x},\mathbf{b}),
\nonumber\\ \label{equality} 
\end{eqnarray} 
where $V_\infty$ is the total volume of the problem (rigorously speaking it can not be bigger than the Born von Karman quantization volume $V_{BK}\rightarrow +\infty$) and  where the surface integral term over the surrounding boundary $\Sigma_\infty=\partial V_\infty$ is given in Appendix A (see Eq.~\ref{A5}). In the DLN approach the surface term vanishes exponentially with the  typical radius $R$ of the surrounding surface. However, here the system is more general and in our Hamiltonian description we are interested in problems where we have vaccum at spatial infinity. Therefore, we should keep this surface term. \\
\indent By keeping the surface contribution in Eq.~\ref{equality}  we can rewrite Eq.~\ref{39c} as:
   \begin{eqnarray}
[\breve{\mathbf{E}}_{\textrm{ret.}}^{(+)(\textrm{eff.})}(\mathbf{a},t_a),\breve{\mathbf{E}}_{\textrm{ret.}}^{(-)(\textrm{eff.})}(\mathbf{b},t_b)]\nonumber\\
=\frac{\hbar}{\pi}\int_0^{+\infty}d\omega\frac{\omega^2}{c^2}\textrm{Imag}[\mathbf{G}_\omega^{(\textrm{eff.})}(\mathbf{a},\mathbf{b})]e^{-i\omega(t_a-t_b)}\nonumber\\
-\frac{\hbar}{\pi}\int_0^{+\infty}d\omega\frac{\omega^2}{c^2}\oint_{\Sigma_\infty}dS\mathbf{F}_\omega^{(\textrm{eff.})}(\mathbf{x},\mathbf{a},\mathbf{b})e^{-i\omega(t_a-t_b)}.
\label{39cc} 
\end{eqnarray} 
In particular if like in the DLN approach the surface term cancels we have  
\begin{eqnarray}
[\breve{\mathbf{E}}_{\textrm{ret.}}^{(+)(\textrm{eff.})}(\mathbf{a},t_a),\breve{\mathbf{E}}_{\textrm{ret.}}^{(-)(\textrm{eff.})}(\mathbf{b},t_b)]\nonumber\\
=\frac{\hbar}{\pi}\int_0^{+\infty}d\omega\frac{\omega^2}{c^2}\textrm{Imag}[\mathbf{G}_\omega^{(\textrm{eff.})}(\mathbf{a},\mathbf{b})]e^{-i\omega(t_a-t_b)}.\nonumber\\
\label{39d} 
\end{eqnarray} 
\indent We see that Eq.~\ref{39d} is very similar to the pure photonic result for vacuum as given by Eq.~\ref{39}. Furthermore, Eq.~\ref{39d} associated with fluctuating currents apparently reduces to Eq.~\ref{39}, i.e., to the result obtained with the pure photon fluctuations, when the local permittivity  $\varepsilon_\omega(\mathbf{x})$ reduces everywhere to $1+i0^+$. For this reason  it is often claimed that the standard DLN formalism without photon fields contains as the limit case the vacuum QED regime. This is interesting and a bit paradoxical  since different origins for fluctuations actually seems to imply an identical result.  The problem is that if the medium is such that $\varepsilon_\omega(\mathbf{x})\rightarrow 1+i0^+$  then the surface integral in Eq.~\ref{equality} does not cancel anymore. Indeed, since  $\varepsilon''_{\omega}(\mathbf{x})\rightarrow 0^+$ we have $\textbf{N}_\omega^{(\textrm{eff.})}(\mathbf{a},\mathbf{b})=\textrm{Imag}[\mathbf{G}_\omega^{(\textrm{eff.})}(\mathbf{a},\mathbf{b})]-\oint_{\Sigma_\infty}dS\mathbf{F}_\omega^{(\textrm{eff.})}(\mathbf{x},\mathbf{a},\mathbf{b})=0$ which in turns implies 
$[\breve{\mathbf{E}}_{\textrm{ret.}}^{(+)(\textrm{eff.})}(\mathbf{a},t_a),\breve{\mathbf{E}}_{\textrm{ret.}}^{(-)(\textrm{eff.})}(\mathbf{b},t_b)]=0$.\\
\indent Therefore, this means that in the vacuum all contributions of the dipole distribution $\widetilde{\mathbf{P}}^{(0)}_\omega$ vanish and the commutator reduces to Eq.~\ref{39} which includes only contributions of the free space photon modes as it should be. In other words, the passage from Eq.~\ref{39c} to Eq.~\ref{39d} is forbidden in vacuum and there is apparently a contradiction with the standard DLN deduction. However, the problem is solved in the QED framework if we remember (see Eq.~\ref{3new}) that in the derivation of the usual DLN approach the term  $[\breve{\mathbf{E}}_{\textrm{in}}^{(+)(\textrm{eff.})}(\mathbf{a},t_a),\breve{\mathbf{E}}_{\textrm{in}}^{(-)(\textrm{eff.})}(\mathbf{b},t_b)]$ in Eq.~\ref{37} cancels since all free modes are infinitely damped by the presence of the residual bulk permittivity~\cite{B} (see Sec. II B). Hence, it is actually the total field commutator $[\breve{\mathbf{E}}^{(+)}(\mathbf{a},t_a),\breve{\mathbf{E}}^{(-)}(\mathbf{b},t_b)]$ that should be written in Eq.~\ref{39d} for the DLN approach:
\begin{eqnarray}
[\breve{\mathbf{E}}^{(+)}(\mathbf{a},t_a),\breve{\mathbf{E}}^{(-)}(\mathbf{b},t_b)]\nonumber\\
=\frac{\hbar}{\pi}\int_0^{+\infty}d\omega\frac{\omega^2}{c^2}\textrm{Imag}[\mathbf{G}_\omega^{(\textrm{eff.})}(\mathbf{a},\mathbf{b})]e^{-i\omega(t_a-t_b)}.\nonumber\\
\label{39e} 
\end{eqnarray}
\indent This discussion shows that at least in the limit  $\varepsilon_{\omega}(\mathbf{x})\rightarrow 1+i 0^+$ both formalisms lead to the same result if we accept to reintroduce the term $[\breve{\mathbf{E}}_{\textrm{in}}^{(+)(\textrm{eff.})}(\mathbf{a},t_a),\breve{\mathbf{E}}_{\textrm{in}}^{(-)(\textrm{eff.})}(\mathbf{b},t_b)]$ which was canceled in the standard DLN approach. Mathematically speaking,  we  have here two ways of taking  the limit.  Either i) we took first  the  limit $\varepsilon_\omega(\mathbf{x})\rightarrow 1+i0^+$  and then afterward we impose $V\rightarrow V_\infty$ or ii) we first fix $\varepsilon_\omega(\mathbf{x})$ then use the geometrical limit  $V\rightarrow V_\infty$ and finally impose $\varepsilon_\omega(\mathbf{x})\rightarrow 1+i0^+$.   The choice i) leads to  an interpretation in term of photon vacuum $[\breve{\mathbf{E}}_{\textrm{in}}^{(+)(v)}(\mathbf{a},t_a),\breve{\mathbf{E}}_{\textrm{in}}^{(-)(v)}(\mathbf{b},t_b)]$ while ii) implies an interpretation in term of material fluctuations (see Eq.~\ref{39e}) i.e.,  with  the idea that the reaction of the bulk medium cancels the field $\breve{\mathbf{E}}_{\textrm{in}}^{(+)(\textrm{eff.})}(\mathbf{x},t)$. Both limiting sequences are thus rigorously equivalent in QED-QNP based on an Hamiltonian treatment.\\
\indent However, the fundamental question is still to know if Eq.~\ref{39e} obtained within the standard DLN model is general and can apply to the case considered in Fig.~1(B) where an inhomogeneous system of local permittivity $\varepsilon_{\omega}(\mathbf{x})$ is surrounded by vacuum. If we return to the difference of structure between the DLN and the usual Huttner Barnett approach (compare Secs. II A and II B) we have apparently some reasons to doubt of  the generality of Eq.~\ref{39e}. Indeed, following the Hamiltonian description summarized in Sec. II A \cite{A,B,C} the QED formalism require both photonic and material degrees of freedom on an equal footing. Therefore, from QED one expects that  the total field commutator  $[\breve{\mathbf{E}}^{(+)}(\mathbf{a},t_a),\breve{\mathbf{E}}^{(-)}(\mathbf{b},t_b)]$  necessarily includes both Eq.~\ref{37} for the photon scattered in the environment and Eq.~\ref{39c} for the  dipole distribution $\widetilde{\mathbf{P}}^{(0)}_\omega(\mathbf{x})$ inside the medium.\\
\indent  Moreover, in agreement with the equivalence theorem obtained in Sec. III, the pure photon field can always be mimicked using a dipole distribution $\widetilde{\mathbf{P}}^{(0)}_\omega(\mathbf{x})$ located in the far field  of the system of interest (i.e. beyond the surface $\Sigma_2$). In this alternative description the field  $\widetilde{\mathbf{E}}_{\textrm{in},\omega}^{(\textrm{eff.,1+2})}(\mathbf{x})\rightarrow 0$ due to the presence of the surrounding absorbing medium of permittivity  $\varepsilon_\omega^{(1)}(\mathbf{x})$. We introduce instead  a new field component $\widetilde{\mathbf{E}}_{\textrm{in},\omega}^{(\textrm{eff.,1+2,'})}(\mathbf{x})$ (see Eq.~\ref{12new}) which is formally equivalent for all practical needs to the scattered photon field $\widetilde{\mathbf{E}}_{\textrm{in},\omega}^{(\textrm{eff.,3})}(\mathbf{x})\rightarrow 0$ (see Eq.~\ref{12c}).\\ 
\indent Therefore, we have now two equivalent ways to write the commutator $[\breve{\mathbf{E}}^{(+)}(\mathbf{a},t_a),\breve{\mathbf{E}}^{(-)}(\mathbf{b},t_b)]$. In the first approach considered previously we have
\begin{eqnarray}
[\breve{\mathbf{E}}^{(+)}(\mathbf{a},t_a),\breve{\mathbf{E}}^{(-)}(\mathbf{b},t_b)]\nonumber\\
=\sum_{\alpha,j}\mathbf{E}_{\alpha,j}^{(\textrm{eff.},3)}(\mathbf{a})\otimes\mathbf{E}_{\alpha,j}^{\ast(\textrm{eff.},3)}(\mathbf{b})e^{-i\omega_\alpha(t_a-t_b)}\nonumber\\
+\frac{\hbar}{\pi}\int_0^{+\infty}d\omega\frac{\omega^2}{c^2}\textbf{N}_\omega^{(\textrm{eff.},3)}(\mathbf{a},\mathbf{b})e^{-i\omega(t_a-t_b)},
\label{39f} 
\end{eqnarray}
 with \begin{eqnarray}
\textbf{N}_\omega^{(\textrm{eff.},3)}(\mathbf{a},\mathbf{b})=\int_{V_2} d^3\mathbf{x}\frac{\omega^2}{c^2}\varepsilon_\omega^{'',(3)}(\mathbf{x})\mathbf{G}_\omega^{(\textrm{eff.})}(\mathbf{a},\mathbf{x})\nonumber\\ \cdot\mathbf{G}_\omega^{\ast,(\textrm{eff.})}(\mathbf{x},\mathbf{b}).\label{mattb}
\end{eqnarray}  Here the label $3$ is to remind that the considered medium is located in the volume $V_2$ which is here surrounded by vacuum like in Fig.~1(A).\\
\indent In the second approach using effective photons we write instead
  \begin{eqnarray}
[\breve{\mathbf{E}}^{(+)}(\mathbf{a},t_a),\breve{\mathbf{E}}^{(-)}(\mathbf{b},t_b)]\nonumber\\
=\frac{\hbar}{\pi}\int_0^{+\infty}d\omega\frac{\omega^2}{c^2}\textbf{N}_\omega^{(\textrm{eff.},3,')}(\mathbf{a},\mathbf{b})e^{-i\omega(t_a-t_b)}\nonumber\\
+\frac{\hbar}{\pi}\int_0^{+\infty}d\omega\frac{\omega^2}{c^2}\textbf{N}_\omega^{(\textrm{eff.},3)}(\mathbf{a},\mathbf{b})e^{-i\omega(t_a-t_b)},
\label{39g} 
\end{eqnarray}  with 
\begin{eqnarray}
\textbf{N}_\omega^{(\textrm{eff.},3,')}(\mathbf{a},\mathbf{b})=\int_{V_1-V_2} d^3\mathbf{x}\frac{\omega^2}{c^2}\varepsilon_\omega^{'',(1)}(\mathbf{x})\mathbf{G}_\omega^{(\textrm{eff.})}(\mathbf{a},\mathbf{x})\nonumber\\ \cdot\mathbf{G}_\omega^{\ast,(\textrm{eff.})}(\mathbf{x},\mathbf{b}),\label{mattc}
\end{eqnarray} where the integral is done over the spatial region $V_1-V_2$. We do not have here to introduce a term $\sum_{\alpha,j}\mathbf{E}_{\alpha,j}^{(\textrm{eff.},1+2)}(\mathbf{a})\otimes\mathbf{E}_{\alpha,j}^{\ast(\textrm{eff.},1+2)}(\mathbf{b})e^{-i\omega_\alpha(t_a-t_b)}$ since for $\mathbf{a},\mathbf{b}\in V_2$ (i.e., far away from $\Sigma_2$) we have $\mathbf{E}_{\alpha,j}^{(\textrm{eff.},1+2)}\approx 0$. Additionally, in this second but equivalent description  we have (i.e. for the same points $\mathbf{a},\mathbf{b}\in V_2$ as previously) the identity    
\begin{eqnarray}
\frac{\hbar}{\pi}\int_0^{+\infty}d\omega\frac{\omega^2}{c^2}\textbf{N}_\omega^{(\textrm{eff.},3,')}(\mathbf{a},\mathbf{b})e^{-i\omega(t_a-t_b)}\nonumber\\
:=\sum_{\alpha,j}\mathbf{E}_{\alpha,j}^{(\textrm{eff.},3)}(\mathbf{a})\otimes\mathbf{E}_{\alpha,j}^{\ast(\textrm{eff.},3)}(\mathbf{b})e^{-i\omega_\alpha(t_a-t_b)},
\label{54} 
\end{eqnarray} which means that the pure photon field commutator of Eq.~\ref{39f} is now completely described by a fluctuating current term over the volume $V_1-V_2$ in agreement with results given in Sec.  III.\\
\indent  Now we do not have to calculate the various complicated terms present in Eq.~\ref{39f} (which includes both photon and matter contributions) or equivalently in Eq.~\ref{39g} (which splits the material contribution into two parts). Indeed, what is relevant is not $\textbf{N}_\omega^{(\textrm{eff.},3,')}(\mathbf{a},\mathbf{b})$  or $\textbf{N}_\omega^{(\textrm{eff.},3)}(\mathbf{a},\mathbf{b})$ but their sum, which reads 
   \begin{eqnarray}
\textbf{N}_\omega^{(\textrm{eff.},1+2)}(\mathbf{a},\mathbf{b})=\int_{V_1} d^3\mathbf{x}\frac{\omega^2}{c^2}\varepsilon_\omega^{'',(1+2)}(\mathbf{x})\mathbf{G}_\omega^{(\textrm{eff.})}(\mathbf{a},\mathbf{x})\nonumber\\ \cdot\mathbf{G}_\omega^{\ast,(\textrm{eff.})}(\mathbf{x},\mathbf{b}).\label{55}
\end{eqnarray} However, from Eq.~\ref{equality} we see that Eq.~\ref{55} can be evaluated if we can compute the surface integral $\oint_{\Sigma_\infty}dS\mathbf{F}_\omega^{(\textrm{eff.},1+2)}(\mathbf{x},\mathbf{a},\mathbf{b})$, where the surface $\Sigma_\infty$ surrounds $V_1$. As shown in Appendix A this integral relies on the knowledge of  the Green tensor $\mathbf{G}_\omega^{(\textrm{eff.},1+2)}(\mathbf{x},\mathbf{a})$ and  $\mathbf{G}_\omega^{(\textrm{eff.},1+2)}(\mathbf{x},\mathbf{b})$  for any points $\mathbf{x}$ on the surface $\Sigma_\infty$ and for $\mathbf{a},\mathbf{b}\in V_2$. This Green  tensor must however cancel since the absorbing media of permittivity $\varepsilon_\omega^{(1)}(\mathbf{x})$ kills any outward propagation at infinity (i.e., like in the standard DLN approach). Therefore, we finally have  from the properties of the Green tensor 
 \begin{eqnarray}
[\breve{\mathbf{E}}^{(+)}(\mathbf{a},t_a),\breve{\mathbf{E}}^{(-)}(\mathbf{b},t_b)]\nonumber\\
=\frac{\hbar}{\pi}\int_0^{+\infty}d\omega\frac{\omega^2}{c^2}\textrm{Imag}[\mathbf{G}_\omega^{(\textrm{eff.},3)}(\mathbf{a},\mathbf{b})]e^{-i\omega(t_a-t_b)},\nonumber\\
\label{39ebis} 
\end{eqnarray}
which is equivalent  to Eq.~\ref{39e} for the points $\mathbf{a},\mathbf{b}\in V_2$ considered (and only for those points).\\
\indent To conclude this calculation we showed that the new DLN description including effective photons leads for points $\mathbf{a},\mathbf{b}\in V_2$ far apart from the boundary $\Sigma_2$ to results similar to those obtained previously within the standard DLN approach. Since this new DLN description is equivalent in practice to the generalized Huttner-Barnett framework used in Sec. II A, and which includes pure photons, we have here a complete QED framework which will for all practical needs be identical to the  former DLN description, but will at once preserve time symmetry and unitarity.\\
\subsection{Some important consequences: spontaneous emission, fluctuations and Casimir forces}
\indent The deductions obtained in the present work will have an impact in many fields of QED and QNP involving fluctuational radiations and sources and this for practical calculations and physical interpretations. This is the case for example when we consider spontaneous emission by a dipolar quantum emitter such as a two-level system located near a nano-antenna. We showed in \cite{C} using the Wigner-Weisskopf approach and the generalized Huttner-Barnett formalism \cite{A,B} how the spontaneous emission rate $\Gamma$ and the local density of states (LDOS) $\rho_{LDOS}(\mathbf{x}_0)$ change with the environment and the position  $\textbf{x}_0$ of the  dipole source.  We have in agreement with the literature~\cite{Novotny}:  
\begin{eqnarray}
\Gamma=\frac{\pi}{3}\frac{\omega_{0}}{\hbar}|\boldsymbol{\mu}_{1,2}|^2\rho_{LDOS}(\mathbf{x}_0),\label{gammab}
\end{eqnarray} 
and 
	\begin{eqnarray}
\rho_{LDOS}(\mathbf{x}_0)=\frac{6\omega_{0}}{\pi c^2}\textrm{Im}[\hat{\textbf{n}}^\ast\cdot\mathbf{G}_\omega^{(\textrm{eff.})}(\mathbf{x}_0,\mathbf{x}_0,\omega_{0}+i0^+)\cdot\hat{\textbf{n}}\label{LDOS} \nonumber\\
\end{eqnarray} with $\boldsymbol{\mu}_{1,2}=|\boldsymbol{\mu}_{1,2}|\hat{\textbf{n}}$ the transition dipole amplitude and $\omega_0$ the transition pulsation.
This results was obtained using the full Hamiltonian including both photonic and material  oscillator contributions. Still, what is remarkable is that it is rigorously identical to the result obtained in classical or semi-classical electrodynamics involving a self interaction field but not zero-point field (zpf) or vacuum fluctuations \cite{Novotny}. Indeed, Eq.~\ref{LDOS} depends on the Green tensor calculated at the position of the source $\textbf{x}_0$, a fact that is reminiscent of the self interaction and field associated with the oscillating dipole. This result is naturally obtained in the standard DLN approach~\cite{Dung2000,Chen2013} and therefore constitutes another illustration of the powerfulness of the DLN methodology (see refs.\cite{Dzotjan2010,Cano2011,Hummer2013,Chen2013,Delga2014,Hakami2014,Choquette2012,Grimsmo2013,Rousseaux2016} for more on this topics in connection with Bloch equations and the DLN formalism).\\ 
\indent Moreover, in the present article we showed how to give a clean foundation to the DLN approach by including dipolar sources located far away from the dipole $\boldsymbol{\mu}_{1,2}$ and its local environment and acting effectively as the pure photon field required in the  generalized Huttner-Barnett formalism~\cite{A,B} (see also \cite{Huttner1992c}). It is not difficult to redo the calculation of \cite{C} with this new method (i.e. without the `real' photon field $\widetilde{\mathbf{E}}_{\textrm{in},\omega}^{(\textrm{eff.,1+2})}(\mathbf{x})\rightarrow 0$ but instead by including the effective photon field $\widetilde{\mathbf{E}}_{\textrm{in},\omega}^{(\textrm{eff.,1+2,'})}(\mathbf{x})$ of dipolar origin) and then to recover Eq.~\ref{LDOS}. This will thus be in complete agreement with the DLN philosophy, which involves only the Green tensor as a fundamental propagative field and the operator $\mathbf{f}^{(0)}_{\omega}(\mathbf{x},t_0)$, $\mathbf{f}^{\dagger(0)}_{\omega}(\mathbf{x},t_0)$ as potential sources of quantum noise.\\
\indent The fundamental commutator Eq.~\ref{39ebis} plays also a key role for the calculation of fluctuations and correlations~\cite{Fermani2006} at different spatial positions and for evaluation of Casimir and thermal forces~\cite{Tomas2002,Buhman2004,Intravaia2014,Intravaia2016,PhilbinCasimir}. Here, within the new DLN formalism the calculations will become more transparent.\\ 
\indent Consider 	as an illustration that the full quantum system (i.e. including pure photonic and material degrees of freedom) is in thermal equilibrium at the temperature $T$. We first observe that since in the region $V_2$ of Fig.~2 the pure photon field $\widetilde{\mathbf{E}}_{\textrm{in},\omega}^{(\textrm{eff.,1+2})}(\mathbf{x})\rightarrow 0$ is absorbed and irrelevant it is only necessary to consider the role of material fluctuations on the Planck formula. More precisely, in agreement with the DLN formalism  the Planck spectrum for the material fluctuating dipoles $\widetilde{\mathbf{P}}^{(0)}_\omega(\mathbf{x})$ leads by definition to~\cite{Fermani2006}: 
\begin{eqnarray}
\langle\mathbf{f}^{\dagger(0)}_{\omega}(\mathbf{x},t)\otimes\mathbf{f}^{(0)}_{\omega'}(\mathbf{x'},t)\rangle_{\textrm{ther.}}=\frac{\delta(\omega-\omega')\delta^3(\mathbf{x}-\mathbf{x'})\textbf{I}}{e^{\frac{\hbar\omega}{k_B T}}-1},\label{tic}
\end{eqnarray}  (where the quantum average $\langle[...]\rangle_{\textrm{ther.}}$ is taken over the Planck distribution) and thus from the canonical commutation~\cite{A} relation to 
\begin{eqnarray}
\langle\mathbf{f}^{(0)}_{\omega}(\mathbf{x},t)\otimes\mathbf{f}^{\dagger(0)}_{\omega'}(\mathbf{x'},t)\rangle_{\textrm{ther.}}=\frac{\delta(\omega-\omega')\delta^3(\mathbf{x}-\mathbf{x'})\textbf{I}}{1-e^{-\frac{\hbar\omega}{k_B T}}},\label{tac}
\end{eqnarray}
where we used $\frac{1}{e^{\frac{\hbar\omega}{k_B T}}-1}+1=\frac{1}{1-e^{-\frac{\hbar\omega}{k_B T}}}$ with $k_B$ the Boltzmann constant.
Now, from  Eqs.~\ref{tic}, \ref{tac}  and by using relations similar to Eq.~\ref{39ebis} for the field correlator we get immediately
\begin{eqnarray}
\langle\breve{\mathbf{E}}^{(-)}(\mathbf{a},t_a)\otimes\breve{\mathbf{E}}^{(+)}(\mathbf{b},t_b)\rangle_{\textrm{ther.}}\nonumber\\
=\frac{\hbar}{\pi}\int_0^{+\infty}d\omega\frac{\omega^2}{c^2}\frac{\textrm{Imag}[\mathbf{G}_\omega^{(\textrm{eff.},3)}(\mathbf{a},\mathbf{b})]}{e^{\frac{\hbar\omega}{k_B T}}-1}e^{-i\omega(t_a-t_b)},\nonumber\\
\label{fluctuatmergitur} 
\end{eqnarray} and
 \begin{eqnarray}
\langle\breve{\mathbf{E}}^{(+)}(\mathbf{a},t_a)\otimes\breve{\mathbf{E}}^{(-)}(\mathbf{b},t_b)\rangle_{\textrm{ther.}}\nonumber\\
=\frac{\hbar}{\pi}\int_0^{+\infty}d\omega\frac{\omega^2}{c^2}\frac{\textrm{Imag}[\mathbf{G}_\omega^{(\textrm{eff.},3)}(\mathbf{a},\mathbf{b})]}{1-e^{-\frac{\hbar\omega}{k_B T}}}e^{-i\omega(t_a-t_b)}.\nonumber\\
\label{fluctuatmergiturb} 
\end{eqnarray} This leads to the total field correlator $\langle\mathbf{E}(\mathbf{a},t_a)\otimes\mathbf{E}(\mathbf{b},t_b)\rangle_{\textrm{ther.}}$ sum of Eq.~\ref{fluctuatmergitur} and Eq.~\ref{fluctuatmergiturb}:
  \begin{eqnarray}
\langle\mathbf{E}(\mathbf{a},t_a)\otimes\mathbf{E}(\mathbf{b},t_b)\rangle_{\textrm{ther.}}\nonumber\\
=\frac{\hbar}{\pi}\int_0^{+\infty}d\omega\frac{\omega^2}{c^2}\textrm{Imag}[\mathbf{G}_\omega^{(\textrm{eff.},3)}(\mathbf{a},\mathbf{b})]\textrm{coth}(\frac{\hbar\omega}{2k_B T}),\nonumber\\
\label{fluctuatmergiturc} 
\end{eqnarray}    
 where we used $\frac{2}{e^{\frac{\hbar\omega}{k_B T}}-1}+1=\frac{1}{1-e^{-\frac{\hbar\omega}{k_B T}}}+\frac{1}{e^{\frac{\hbar\omega}{k_B T}}-1}=\textrm{coth}(\frac{\hbar\omega}{2k_B T})$. This is a purely quantum formulation of the fluctuation dissipation theorem agreeing with both the standard DLN approach  and the much older phenomenological noise formulation proposed by Rytov and Lifshitz~\cite{Lifshitz1956,Ginzburg,Rytov,Milonnibook,Novotny} (i.e., extensively used in the recent years in the field of `fluctuational electrodynamics' for interpreting Casimir and thermal forces at the nanoscale~\cite{Novotny,Rosa2010,Agarwal1975,Sipe1984,Rousseau,Henkel,Otey2014}). Importantly, this result reduces to $-i\hbar\boldsymbol{\Delta}_{\textrm{ret.}}^{(+)(v)}(|t_a-t_b|,\mathbf{a},\mathbf{b})$ in the vacuum case and has an usual interpretation as the retarded field propagator~\cite{QM}.\\
\indent Now, in the new DLN formulation we can compute the fluctuational force acting on a body and resulting from the thermal bath considered before. For this we use the standard dipolar force formula derived in \cite{A,Novotny}  and which reads 
\begin{eqnarray}
\langle\mathbf{F}(t)\rangle=\int_{\delta V} d^3\mathbf{ x}\sum_i \langle P_i(\mathbf{x},t)\boldsymbol{\nabla}E_i(\mathbf{x},t)\rangle,
\end{eqnarray} 
with $P_i(\textbf{x},t)$ the $i^{th}$ component ($i=1,2,3$) of the total dipole density distribution in the body of volume  $\delta V<< V_2$ and $E_i(\textbf{x},t)$ is the total electric field operator acting upon this dipole distribution. We remind that this expression for the force is rigorously valid  only in the quasi-static limit when the role of motion and  magnetic field can be neglected~\cite{A,Novotny}. In the case of the thermal distribution considered previously we get after some calculations summarized in Appendix B the total thermal-Casimir static force acting upon the body:  
 \begin{eqnarray}\langle\mathbf{F}\rangle_{\textrm{ther.}}=\int_{\delta V} d^3\mathbf{ x}\frac{\hbar}{\pi}\int_0^{+\infty}d\omega\frac{\omega^2}{c^2}\textrm{coth}(\frac{\hbar\omega}{2k_B T})\nonumber\\
\cdot\textrm{Imag}[(\varepsilon_\omega^{(3)}(\textbf{x})-1)\boldsymbol{\nabla}_1\textrm{Tr}[\mathbf{G}_\omega^{(\textrm{eff.},3)}(\mathbf{x},\mathbf{x})]]\nonumber\\
=\int_{\delta V} d^3\mathbf{ x}\frac{\hbar}{\pi}\int_{-\infty}^{+\infty}d\omega\frac{\omega^2}{c^2}\frac{\textrm{Imag}[(\varepsilon_\omega^{(3)}(\textbf{x})-1)}{1-e^{-\frac{\hbar\omega}{k_B T}}}\nonumber\\ \cdot \boldsymbol{\nabla}_1 \textrm{Tr}[\mathbf{G}_\omega^{(\textrm{eff.},3)}(\mathbf{x},\mathbf{x})]]\label{Casimir}
\end{eqnarray}
  where $Tr[...]$ is the trace operator and $\boldsymbol{\nabla}_1$ is a gradient operator acting only on the left $\textbf{x}$ variable in $\mathbf{G}_\omega^{(\textrm{eff.},3)}(\mathbf{x},\mathbf{x})$. \\ \indent Remarkably, this formula is rigorously identical to the expression obtained in the `fluctuational electrodynamical' framework \cite{Novotny,Rosa2010,Agarwal1975,Sipe1984,Rousseau,Henkel,Otey2014}. Here it is obtained within the new DLN formalism which includes effective photons and which is equivalent (as we showed in Sec. III) to the generalized Huttner-Barnett formalism developed in \cite{A,B,C}.\\ 
\indent Moreover, this is crucial here concerning the debate about the physical origin  of the Casimir force~\cite{Milonnibook}. Indeed, in the generalized Huttner-Barnett formalism  we have pure photonic and dipolar fluctuations at work. Both are mandatory in this Hamiltonian approach for interpreting the Casimir force and at the same time in order to respect the complete unitarity and time symmetry of the Hamiltonian dynamics. However, from the equivalence theorem demonstrated in the present work we now have the possibility to interpret the  Casimir force only as resulting of dipole fluctuations. But, in order to do that, we have not only have to include dipoles located in the material body considered (here of volume $\delta V$ in Eq.~\ref{Casimir}) but also dipoles located in the far-field (i.e. in the region $V_1-V_2$) and emitting a field acting as effective photons. However, like for the LDOS formula in Eq.~\ref{LDOS} the results in Eq.~\ref{Casimir} only depends on local properties in the region of the body (i.e. $\delta V$). Therefore, at the end everything is identical to the result obtained within the old DLN formalism without the pure photon field (see Fig.~1(B)) and without effective photon field (compare with Fig.~2).  As we reminded  before the DLN approach has an old history and was already used by Lifshitz and Rytov in order to justify the Casimir force formula~\cite{Lifshitz1956,Ginzburg,Rytov} and later it was naturally used in the quantized version of the DLN~\cite{Tomas2002,Buhman2004,Intravaia2014,Intravaia2016,PhilbinCasimir}. The standard DLN approach apparently differs strongly in essence from the so-called scattering approach~\cite{Casimirpaper,Jaekel1991,Reynaud2000} that considers the radiation pressure exerted by scattered optical modes on the material system. The scattering approach considers therefore only pure photon modes, i.e. the role of zpf for light, and was originally developed for lossless and consequently noncausal dielectric systems. It is possible to extend the scattering formalism by including  some additional propagation channels for the photons acting as attenuators~\cite{Jeffers1993,Tame2008,Ballester2009,Ballester2010,Tame2013}, which leads to a causal discussion of the Casimir force in agreement with Kramers-Kronig formula \cite{Genet2003,Genet2004,Lambrecht2006,Ingold2015}. Moreover, the scattering formalism with the supplementary hidden optical modes acting as attenuators is not so different from the Hamiltonian Huttner-Barnett formalism~\cite{Huttner1991,Huttner1992a,Huttner1992b,Huttner1992c,Matloob1995,Matloob1996,Barnett1995,A,B,C} which attributes the origin of loss and dispersion to the coupling of photons to a bath of material harmonic oscillators. Therefore, ultimately all theories are expected to give the same results, e.g., for Casimir and thermal forces calculations. However, in the present work we showed that the DLN approach should be properly generalized by including dipolar sources in the far-field acting as effective photons.  With such modeling of the effective photon field we have demonstrated the equivalence with the generalized Huttner-Barnett approach for inhomogeneous media. Subsequently, we should also have equivalence with the scattering approach if properly generalized (this is however going beyond the present work).                
\section{Summary and perspectives}
\indent To summarize: in this work we compared different theoretical approaches for analyzing QNP and QED in complex inhomogeneous dielectric systems. We started (see Sec. II A) with the generalized Huttner-Barnett Hamiltonian  formulation developed in \cite{A,B,C} which extends to the inhomogeneous medium case the works done in~\cite{Huttner1991,Huttner1992a,Huttner1992b,Huttner1992c,Matloob1995,Matloob1996,Barnett1995} for homogeneous dielectrics. We compared this approach with the DLN method  (see Sec. II B) developed by Gruner and Welsch~\cite{Gruner1995,Gruner1996,Yeung1996}, and which extends the fluctuational electrodynamics developed by Lifshitz, Rytov and others~\cite{Lifshitz1956,Ginzburg,Rytov,Milonnibook,Rosa2010,Agarwal1975,Sipe1984,Rousseau,Henkel,Otey2014}. In the Huttner-Barnett formalism the quantized description requires in general a pure photon field $\mathbf{E}_{\textrm{in}}^{(\textrm{eff.})}(\mathbf{x},t)$ corresponding to vacuum photon modes scattered by the complex dielectric environment.  We should also include in this formalism a retarded source electric field $\mathbf{E}_{\textrm{ret.}}^{(\textrm{eff.})}(\mathbf{x},t)$ emitted by the dipole distribution $\mathbf{P}^{(\textrm{eff.})}(\mathbf{x'},t')$  sum of the molecular dipole distribution  $\mathbf{P}^{(\textrm{mol.})}(\mathbf{x'},t')$ located in the environment and the dielectric dipole distribution $\mathbf{P}^{(0)}(\mathbf{x'},t')$ associated with the material degrees of freedom in the dielectric system itself. In  the DLN approach the pure photon field is missing since it is absorbed by a residual bulk permittivity killing all scattered photon modes coming from infinity. Due to the strong differences between the Huttner Barnett and the DLN methods it was not however clear how to compare the calculations. Since the DLN approach is widely used this is an important issue for QNP. \\
\indent In Sec. III we showed how to construct an effective medium which for all practical calculations demonstrates an equivalence  between the Huttner Barnett formalism and the DLN approach. The idea is to surround the physical system considered by a weakly dissipative dielectric medium located in the far-field (see Fig.~2).  The effect of this surrounding medium is twice. On the one side it absorbs all scattered pure photon modes coming from infinity:  $\mathbf{E}_{\textrm{in}}^{(\textrm{eff.})}(\mathbf{x},t)\rightarrow 0 $, which are therefore inoperative on the physical  system considered. On the other side, the surrounding  medium creates, through its own dielectric dipole distribution $\mathbf{P}^{(0)}(\mathbf{x'},t')$, an effective photon field $\mathbf{E}_{\textrm{in}}^{(\textrm{eff.},')}(\mathbf{x},t) $ having all the physical and mathematical properties of a scattered  photon field $\mathbf{E}_{\textrm{in}}^{(\textrm{eff.})}(\mathbf{x},t)$. Within this alternative DLN formulation we have thus complete equivalence between the DLN and Huttner-Barnett formulations of QED and QNP. Remarkably, the old and new DLN approaches give the same results since the surrounding medium has not effect on local properties inside the physical system considered. We illustrated this fundamental issue with few examples associated with quantum fluctuations such as spontaneous  emission, quantum correlations and Casimir forces at finite temperature. Using the DLN formalism leads to simple analysis determined by the complex Green tensor  and the local permittivity in the system considered and therefore to transparent expressions that agree with the older fluctuational electrodynamics of Rytov.\\ 
\indent We think that the present analysis will motivate further works concerning the links between the different methods used in QED and QNP which play a fundamental role in nano-plasmonics, non linear optics and mechanical motions a the nanoscale using Casimir and thermal forces.

 \section{Acknowledgments}
\indent This work was supported by Agence Nationale de la Recherche (ANR), France, through the PLACORE (ANR-13-BS10-0007) grant. The author gratefully acknowledges a critical reading by S. Huant.  
\appendix
\section{The dyadic-dyadic Green theorem and some relations}
\indent Let $\mathbf{Q}(\mathbf{x})$ and $\mathbf{P}(\mathbf{x})$ two spatially dependent dyads. The dyadic-dyadic Green theorem states that in a volume $V$ surrounded by the surface $\Sigma$ we have     
\begin{eqnarray}
\int_{V}d^3\mathbf{x}\left([\boldsymbol{\nabla}\times\boldsymbol{\nabla}\times\mathbf{Q}]^T\cdot\mathbf{P}-\mathbf{Q}^T\cdot[\boldsymbol{\nabla}\times\boldsymbol{\nabla}\times\mathbf{P}]\right)\nonumber\\
=\oint_{\Sigma}dS\left([\boldsymbol{\nabla}\times\mathbf{Q}]^T\cdot(\hat{\mathbf{n}}\times\mathbf{P})+\mathbf{Q}^T \cdot[\hat{\mathbf{n}}\times\boldsymbol{\nabla}\times\mathbf{P}]\right),\label{A1} 
\end{eqnarray} where $\hat{\mathbf{n}}$ is the outwardly oriented unit vector normal to the surface element $dS$ of $\Sigma$ and $T$  is the transpose operator.\\
\indent Consider first the choice  $\mathbf{Q}(\mathbf{x})=\mathbf{G}_\omega^{(\textrm{eff.})}(\mathbf{x},\mathbf{a})$ and $\mathbf{P}(\mathbf{x})=\mathbf{G}_\omega^{(\textrm{eff.})}(\mathbf{x},\mathbf{b})$
with $\mathbf{a}$, and $\mathbf{b}$ two positions inside the volume $V$. With this choice we will obtain the reciprocity theorem. While this result is well established, we will review it briefly here since its deduction plays a central role in our demonstration.  From Eqs.~\ref{A1} and \ref{Green} we  thus deduce 
\begin{eqnarray}
\mathbf{G}_\omega^{(\textrm{eff.})}(\mathbf{a},\mathbf{b})-\mathbf{G}_\omega^{T,(\textrm{eff.})}(\mathbf{b},\mathbf{a})\nonumber\\=\oint_{\Sigma}dS\left([\boldsymbol{\nabla}\times\mathbf{Q}]^T\cdot(\hat{\mathbf{n}}\times\mathbf{P})+\mathbf{Q}^T \cdot[\hat{\mathbf{n}}\times\boldsymbol{\nabla}\times\mathbf{P}]\right).\label{A2} 
\end{eqnarray}
Now we consider the limit where the surface $\Sigma$ is spherical  with a radius $R\rightarrow+\infty$ and we suppose that at infinity the permittivity approaches a finite value $\varepsilon_{\omega}^{(\textrm{bulk})}$. In this regime, the Green tensor at infinity decays with  $R$ as $\propto\frac{e^{i\omega\sqrt{\varepsilon_{\omega}^{(\textrm{bulk})}}R/c}}{R}$ 
and, if the bulk medium is causal and therefore lossy, it involves an exponential decay that reduces the surface integral in Eq.~\ref{2} to zero.  Therefore we obtain 
\begin{eqnarray}
\mathbf{G}_\omega^{(\textrm{eff.})}(\mathbf{a},\mathbf{b})=\mathbf{G}_\omega^{T,(\textrm{eff.})}(\mathbf{b},\mathbf{a}),\label{recip}
\end{eqnarray}
 which is a statement of Lorentz's reciprocity theorem. However, this result is actually much more robust and does not require having an absorbing media at infinity. Indeed, if this medium is lossless, i.e., if  $\varepsilon_{\omega}^{(\textrm{bulk})}=1$ (which corresponds to vacuum), we can use the Sommerfeld radiation condition for any point $\mathbf{x}$ on the surface $\Sigma$, i.e.,  $\boldsymbol{\nabla}\times\mathbf{G}_\omega^{(\textrm{eff.})}(\mathbf{x},\mathbf{u})\simeq i\frac{\omega\sqrt{\varepsilon_{\omega}^{(\textrm{bulk})}}}{c}\hat{\mathbf{R}}\times\mathbf{G}_\omega^{(\textrm{eff.})}(\mathbf{x},\mathbf{u})$, with $\mathbf{u}=\mathbf{a}$ or $\mathbf{b}$ and $\hat{\mathbf{R}}=\hat{\mathbf{n}}$ the unit radial vector oriented  outwardly to the surrounding sphere. The Sommerfeld condition states that at spatial infinity the radiated field (directed outwardly) has locally a plane wave structure propagating in a medium of permittivity $\varepsilon_{\omega}^{(\textrm{bulk})}$.
Insertion of the Sommerfeld radiation condition in Eq.~\ref{A2} shows that the two terms in the surface integral compensate each other and therefore the reciprocity theorem Eq.~\ref{recip} is valid even if the surrounding medium is actually vacuum.\\
\indent For the present work we now consider a different choice for  $\mathbf{Q}(\mathbf{x})=\mathbf{G}_\omega^{(\textrm{eff.})}(\mathbf{x},\mathbf{a})$ and $\mathbf{P}(\mathbf{x})=\mathbf{G}_\omega^{\ast,(\textrm{eff.})}(\mathbf{x},\mathbf{b})$. With such a choice we obtain similarly as for Eq.~\ref{A2} the relation: 
 \begin{eqnarray}
\mathbf{G}_\omega^{\ast,(\textrm{eff.})}(\mathbf{a},\mathbf{b})-\mathbf{G}_\omega^{T,(\textrm{eff.})}(\mathbf{b},\mathbf{a})\nonumber\\
+2i\int_{V}d^3\mathbf{x}\frac{\omega^2}{c^2}\varepsilon''_{\omega}(\mathbf{x})\mathbf{G}_\omega^{T,(\textrm{eff.})}(\mathbf{x},\mathbf{a})\mathbf{G}_\omega^{\ast,(\textrm{eff.})}(\mathbf{x},\mathbf{b})
\nonumber\\=\oint_{\Sigma}dS\left([\boldsymbol{\nabla}\times\mathbf{Q}]^T\cdot(\hat{\mathbf{n}}\times\mathbf{P})+\mathbf{Q}^T \cdot[\hat{\mathbf{n}}\times\boldsymbol{\nabla}\times\mathbf{P}]\right).\label{A4} 
\end{eqnarray} 
Moreover, by using  the reciprocity theorem and the Sommerfeld radiation condition on a sphere  $\Sigma_\infty$ of radius $R\rightarrow +\infty$ we get Eq.\ref{equality} with 
\begin{eqnarray}
\oint_{\Sigma_\infty}dS\mathbf{F}_\omega^{(\textrm{eff.})}(\mathbf{x},\mathbf{a},\mathbf{b})
=\frac{\omega}{c}\sqrt{\varepsilon_{\omega}^{(\textrm{bulk})}}\oint_{\Sigma_\infty}dS\mathbf{G}_\omega^{T,(\textrm{eff.})}(\mathbf{x},\mathbf{a})\nonumber\\ \cdot[\hat{\mathbf{R}}\times\hat{\mathbf{R}}\times\mathbf{G}_\omega^{\ast,(\textrm{eff.})}(\mathbf{x},\mathbf{b})]. \nonumber\\ \label{A5} 
\end{eqnarray} 
However, contrarily to what occurs for the reciprocity theorem the Sommerfeld radiation condition is not sufficient to eliminate the surface integral. We emphasize that in the DLN approach~\cite{Gruner1995,Gruner1996,Yeung1996,Scheelreview2008} the bulk medium is supposed lossy at spatial infinity and therefore due to the asymptotic decay of the Green tensor as $\propto\frac{e^{i\omega\sqrt{\varepsilon_{\omega}^{(\textrm{bulk})}}R/c}}{R}$ the surface term cancels.
\section{Calculation of Casimir forces within the Langevin noise approach}
\indent We start with the standard dipole expression \cite{Novotny} for the force which was derived within a QED framework in \cite{A}: $\langle\mathbf{F}(t)\rangle=\int_{\delta V} d^3\mathbf{ x}\sum_i \langle P_i(\mathbf{x},t)\boldsymbol{\nabla}E_i(\mathbf{x},t)\rangle$. Here we write $\langle\mathbf{F}(t)\rangle=\langle\mathbf{F}^{(1)}(t)\rangle+\langle\mathbf{F}^{(1)}(t)\rangle$, with 
\begin{eqnarray}
\langle\mathbf{F}^{(1)}(t)\rangle=\int_{\delta V} d^3\mathbf{ x}\sum_i \langle \breve{P}^{(+)}_i(\mathbf{x},t)\boldsymbol{\nabla}\breve{E}^{(-)}_i(\mathbf{x},t)\rangle_{\textrm{ther.}}\nonumber\\
\langle\mathbf{F}^{(2)}(t)\rangle=\int_{\delta V} d^3\mathbf{ x}\sum_i \langle \breve{P}^{(-)}_i(\mathbf{x},t)\boldsymbol{\nabla}\breve{E}^{(+)}_i(\mathbf{x},t)\rangle_{\textrm{ther.}}.\nonumber\\
\end{eqnarray}
In order to calculate these terms we use the definitions
\begin{eqnarray}
\breve{\textbf{P}}^{(+)}(\mathbf{x},t)=\int_0^{+\infty}d\omega[\sqrt{\frac{\hbar\varepsilon''_{\omega}(\mathbf{x})}{\pi}}\mathbf{f}^{(0)}_{\omega}(\mathbf{x},t)\nonumber\\
+(\varepsilon_\omega(\textbf{x})-1)\int d^3 \mathbf{x'}\frac{\omega^2}{c^2}\mathbf{G}_\omega^{(\textrm{eff.})}(\mathbf{x},\mathbf{x'})\nonumber\\ \cdot\sqrt{\frac{\hbar\varepsilon''_{\omega}(\mathbf{x'})}{\pi}}\mathbf{f}^{(0)}_{\omega}(\mathbf{x'},t)],\nonumber\\
\end{eqnarray}
\begin{eqnarray}
\breve{\textbf{E}}^{(+)}(\mathbf{x},t)=\int_0^{+\infty}d\omega\int d^3 \mathbf{x'}\frac{\omega^2}{c^2}\mathbf{G}_\omega^{(\textrm{eff.})}(\mathbf{x},\mathbf{x'})\nonumber\\ \cdot\sqrt{\frac{\hbar\varepsilon''_{\omega}(\mathbf{x'})}{\pi}}\mathbf{f}^{(0)}_{\omega}(\mathbf{x'},t),\nonumber\\
\end{eqnarray} and $\breve{\textbf{P}}^{(-)}(\mathbf{x},t)=(\breve{\textbf{P}}^{(+)}(\mathbf{x},t))^\dagger$, $\breve{\textbf{E}}^{(-)}(\mathbf{x},t)=(\breve{\textbf{E}}^{(+)}(\mathbf{x},t))^\dagger$ 
Using these definitions  and Eq.~\ref{tac} we write the first term as $\langle\mathbf{F}^{(1)}(t)\rangle=\langle\mathbf{F}^{(11)}(t)\rangle+\langle\mathbf{F}^{(12)}(t)\rangle$, with 
\begin{eqnarray}
\langle\mathbf{F}^{(11)}(t)\rangle=\int_{\delta V} d^3\mathbf{ x}\frac{\hbar}{\pi}\int_0^{+\infty}d\omega\frac{\omega^2}{c^2}\varepsilon''_\omega(\textbf{x})\nonumber\\ \cdot\frac{\boldsymbol{\nabla}_1\textrm{Tr}[\mathbf{G}_\omega^{\ast,(\textrm{eff.})}(\mathbf{x},\mathbf{x})]}{1-e^{-\frac{\hbar\omega}{k_B T}}},\nonumber\\
\end{eqnarray}
and
\begin{eqnarray}
\langle\mathbf{F}^{(12)}(t)\rangle=\int_{\delta V} d^3\mathbf{ x}\frac{\hbar}{\pi}\int_0^{+\infty}d\omega\frac{\omega^2}{c^2}\frac{(\varepsilon_\omega(\textbf{x})-1)}{1-e^{-\frac{\hbar\omega}{k_B T}}}\nonumber\\ =\sum_{ij} \int d^3\mathbf{x'}\frac{\omega^2}{c^2}\varepsilon''_{\omega}(\mathbf{x'})G_{\omega,ij}^{(\textrm{eff.})}(\mathbf{x},\mathbf{x'})\boldsymbol{\nabla}_\textbf{x}G_{\omega,ij}^{\ast,(\textrm{eff.})}(\mathbf{x},\mathbf{x'})\nonumber\\
=\int_{\delta V} d^3\mathbf{ x}\frac{\hbar}{\pi}\int_0^{+\infty}d\omega\frac{\omega^2}{c^2}\frac{(\varepsilon_\omega(\textbf{x})-1)}{1-e^{-\frac{\hbar\omega}{k_B T}}}\nonumber\\ \cdot \boldsymbol{\nabla}_1\textrm{Tr}[\textrm{Imag}[\mathbf{G}_\omega^{(\textrm{eff.})}(\mathbf{x},\mathbf{x})]].\nonumber\\
\label{force12}
\end{eqnarray}
In going from the second to the last line of Eq.~\ref{force12} we used some properties of the partial derivative for the Green tensor:\\
\indent First, from the reciprocity theorem, we have $G_{\omega,ij}^{(\textrm{eff.})}(\mathbf{a},\mathbf{b})=G_{\omega,ji}^{(\textrm{eff.})}(\mathbf{b},\mathbf{a})$ and, therefore, we have $\boldsymbol{\nabla}_\textbf{a}G_{\omega,ii}^{(\textrm{eff.})}(\mathbf{a},\mathbf{b})=\boldsymbol{\nabla}_\textbf{a}G_{\omega,ii}^{(\textrm{eff.})}(\mathbf{b},\mathbf{a})$, which implies  
\begin{eqnarray}
\boldsymbol{\nabla}_1 G_{\omega,ii}^{(\textrm{eff.})}(\mathbf{x},\mathbf{x})=\boldsymbol{\nabla}_2 G_{\omega,ii}^{(\textrm{eff.})}(\mathbf{x},\mathbf{x}),\label{symdiv}\end{eqnarray} where $\boldsymbol{\nabla}_1$ (respectively $\boldsymbol{\nabla}_2$) acts on the left (respectively right) \textbf{x} variable of the Green tensor.\\
 \indent Second, from  Eq.~\ref{equality} which is valid for $\textbf{a}$, $\textbf{b}$ near the center of region $V_2$ we have
\begin{eqnarray}
\boldsymbol{\nabla}_\textbf{b}\textrm{Imag}[G_{\omega,ii}^{(\textrm{eff.})}(\mathbf{a},\mathbf{b})]
=\int d^3\mathbf{x}\frac{\omega^2}{c^2}\varepsilon''_{\omega}(\mathbf{x})G_{\omega,ij}^{(\textrm{eff.})}(\mathbf{a},\mathbf{x})\nonumber\\ \cdot \boldsymbol{\nabla}_\textbf{b}G_{\omega,ij}^{\ast,(\textrm{eff.})}(\mathbf{b},\mathbf{x})
\nonumber\\ \label{equalityder} 
\end{eqnarray} which therefore leads to 
\begin{eqnarray}
\boldsymbol{\nabla}_2\textrm{Imag}[G_{\omega,ii}^{(\textrm{eff.})}(\mathbf{x},\mathbf{b})]
=\int d^3\mathbf{x'}\frac{\omega^2}{c^2}\varepsilon''_{\omega}(\mathbf{x'})G_{\omega,ij}^{(\textrm{eff.})}(\mathbf{x},\mathbf{x'})\nonumber\\ \cdot \boldsymbol{\nabla}_\textbf{x}G_{\omega,ij}^{\ast,(\textrm{eff.})}(\mathbf{x},\mathbf{x'}).
\nonumber\\ \label{equalityderder} 
\end{eqnarray} Inserting Eq.~\ref{equalityderder} together with the symmetry given by Eq..~\ref{symdiv} in Eq.~\ref{force12} allows us to justify the last line of this equation. Finally, regrouping $\langle\mathbf{F}^{(11)}(t)\rangle$  and $\langle\mathbf{F}^{(12)}(t)\rangle$ leads directly to 
\begin{eqnarray}
\langle\mathbf{F}^{(1)}(t)\rangle=\int_{\delta V} d^3\mathbf{ x}\frac{\hbar}{\pi}\int_0^{+\infty}d\omega\frac{\omega^2}{c^2}\textrm{Imag}[\frac{(\varepsilon_\omega(\textbf{x})-1)}{1-e^{-\frac{\hbar\omega}{k_B T}}}\nonumber\\ \cdot \boldsymbol{\nabla}_1\textrm{Tr}[\mathbf{G}_\omega^{(\textrm{eff.})}(\mathbf{x},\mathbf{x})]].\nonumber\\
\label{force1}
\end{eqnarray}  
\indent We can do similar calculations for $\langle\mathbf{F}^{(2)}(t)\rangle$ and we get 
\begin{eqnarray}
\langle\mathbf{F}^{(1)}(t)\rangle=\int_{\delta V} d^3\mathbf{ x}\frac{\hbar}{\pi}\int_0^{+\infty}d\omega\frac{\omega^2}{c^2}\textrm{Imag}[\frac{(\varepsilon_\omega(\textbf{x})-1)}{e^{\frac{\hbar\omega}{k_B T}}-1}\nonumber\\ \cdot \boldsymbol{\nabla}_1\textrm{Tr}[\mathbf{G}_\omega^{(\textrm{eff.})}(\mathbf{x},\mathbf{x})]],\nonumber\\
\label{force2}
\end{eqnarray} and, therefore, Eq.~\ref{Casimir}.   

\end{document}